\documentclass[fleqn,usenatbib]{mnras}

\usepackage{newtxtext,newtxmath}

\usepackage[T1]{fontenc}
\usepackage{ae,aecompl}

\usepackage{graphicx}	
\usepackage{amsmath}	
\usepackage{amssymb}	

\title[]{On the robustness of analysis techniques for molecular detections using high resolution exoplanet spectroscopy}

\author[Samuel. H. C. Cabot]{Samuel. H. C. Cabot$^{1}$,
Nikku Madhusudhan$^{1}$\thanks{E-mail: nmadhu@ast.cam.ac.uk (NM)},
George A. Hawker$^{1}$
\newauthor and Siddharth Gandhi$^{1}$
\\
$^{1}$Institute of Astronomy, University of Cambridge, Madingley Road, Cambridge, CB3 0HA, UK
}

\date{Accepted XXX. Received YYY; in original form ZZZ}

\pubyear{2015}

\begin{document}
\label{firstpage}
\pagerange{\pageref{firstpage}--\pageref{lastpage}}
\maketitle

\begin{abstract}

High-resolution doppler spectroscopy provides a powerful means for chemical detections in exoplanetary atmospheres. This approach involves monitoring hundreds of molecular lines in the planetary spectrum doppler shifted by the orbital motion of the planet. The molecules are detected by cross-correlating the observed spectrum of the system with a model planetary spectrum. The method has led to molecular detections of H$_2$O, CO, and TiO in hot Jupiters using large ground-based telescopes. Critical to this method, however, is the accurate removal of the stellar and telluric features from the observed spectrum, also known as detrending. Previous molecular detections have relied on specific choices of detrending methods and parameters. However, the robustness of molecular detections across the different choices has not been investigated in detail. We conduct a systematic investigation of the effect of detrending algorithms, parameters, and optimizations on chemical detections using high-resolution spectroscopy. As a case study, we consider the hot Jupiter HD~189733~b. Using multiple methods, we confirm high-significance detections of H$_2$O (4.8$\sigma$) and CO (4.7$\sigma$). Additionally, we report evidence for HCN at high significance (5.0$\sigma$). On the other hand, our results highlight the need for improved metrics and extended observations for robust confirmations of such detections. In particular, we show that detection significances of $\gtrsim$ 4$\sigma$ can be obtained by optimizing detrending at incorrect locations in the planetary velocity space; such false positives can occur in nearly 30\% of cases. We discuss approaches to help distinguish molecular detections from spurious noise. 
\end{abstract}




\section{Introduction}

Detections of chemical species in the atmospheres of exoplanets bring us closer to finding Earth-like worlds and signs of life beyond our Solar System. Chemical species can provide important insights into the atmospheric processes and  formation pathways of exoplanets \citep{madhu2016}. Despite the discovery of thousands of exoplanets, atmospheric characterization remains a difficult task. Attempts to identify chemical species and their abundances in exoplanet atmospheres are often limited by large uncertainties or degeneracies. However, a recent approach to atmospheric characterization using high-resolution Doppler spectroscopy \citep{snellen2010} has led to high-confidence detections of molecules in the atmospheres of hot Jupiters, highly-irradiated gas giant planets with no analogue in our Solar System. 

High-resolution spectroscopy has been used with time-series transmission or thermal emission spectra to directly measure the Doppler shift of planetary spectral lines. Recent studies have made high-fidelity detections of molecules in the near-infrared (NIR) with the CRIRES spectrograph (R$\sim 100,000$) at the {\it VLT} \citep{birkby2018}. Additionally, chemical detections have been made with extended observations using GIANO at {\it Telescopio Nazionale Galileo} \citep{brogi2018}, HDS at {\it Subaru} \citep{nugroho2017}, and NIRSPEC at {\it Keck} \citep{piskorz2016}. Inferences of chemical species include CO and H$_2$O in transmission spectra \citep{snellen2010,brogi2016} and CO, H$_2$O, TiO, and HCN \citep{brogi2012,birkby2013,rodler2013,nugroho2017,hawker2018} in emission spectra. Detections are usually made by cross-correlating a model template spectrum with the data, based on a dense forest of thousands of individual absorption lines. 

A central aspect of molecular detections using this method is accurate removal of extraneous signals in a process known as detrending. It is imperative to remove the dominant stellar spectrum and telluric absorption by Earth's atmosphere, in addition to systematics in the data. These contributions would otherwise affect the cross-correlation of the model template with the intrinsic planetary spectral features. This detrending process is at the crux of the high-resolution spectroscopic analyses. Detrending algorithms exploit the fact that planetary features are Doppler shifted by tens of km s$^{-1}$ over the course of observations, corresponding to tens of pixels on the detector chip. However, stellar, telluric and instrumental effects remain approximately constant with time since their sources are nearly stationary. Detrending involves modeling and removing these nuisance signals, while minimally degrading the planetary signal. 

Detrending is in itself a complex task, and previous studies used different methods to address it. One non-parametric method uses functional fits to the measured atmospheric airmass to remove low-order trends, and sampling of residuals in the data to remove higher-order trends \citep{snellen2010,brogi2012}. Other methods include low-rank approximations to the data \citep{dekok2013,birkby2013,birkby2017}, and direct modeling of telluric features \citep{lockwood2014}. These latter detrending methods can involve fine-tuning several parameters relating to masking data which was corrupted by either instrumental effects or severe telluric contamination, or to setting aggressiveness of the detrending routine (e.g. how many underlying trends to model, or whether to use a high-pass filter).

When sampling residuals from the data or determining the values of detrending parameters, previous studies strike a balance between maximizing the detection of the planetary signal and maintaining a stable and consistent detection. Small changes to parameters, removing additional trends, or sampling different residuals do not significantly change the detection significance. Since the data is highly varied, and there are unknown high-order systematics, it is difficult to analytically identify the trends themselves or the appropriate detrending parameters. Detrending is usually optimized by injecting an artificial planet signal into the data, and maximizing the significance of its recovery \citep{birkby2013,birkby2017,nugroho2017}. However, there is no consistent treatment of all of detrending nuances in the literature as noted in \citet{hawker2018}. 

In this study, we thoroughly investigate how molecular detection significances depend on the underlying detrending process. We first select the two the most common detrending algorithms and demonstrate the robustness of our pipeline by confirming high-significance detections of CO in Tau Boo b \citep{brogi2012} and H$_2$O in HD 189733 b \citep{birkby2013}. We then vary the detrending hyperparameters, and subsequently measure the variation in detection significances of molecular species. These dependences are extremely important in determining how susceptible detrending is to false-positives. As a test case, we analyze dayside observations of the hot Jupiter HD 189733 b. The planet orbits a nearby bright K0V star with magnitude V = 7.7 making it a good candidate for high-resolution atmospheric spectroscopy \citep{2011IAUS..276..208S}. The orbital period of the planet is 2.2 days giving it an equilibrium temperature of $\sim$1200 K which is conducive for the presence of several key molecules in the atmosphere such as H$_2$O, CO and HCN \citep{madhu2012,moses2013}. 

Our study allows us to make molecular detections as well as quantify the risks associated with false-positives. We discuss the implementation of our high-resolution spectral analysis pipeline in Section~\ref{sec2}. We use different detrending methods to confirm high-significance detections of H$_2$O and CO from previous studies in Section~\ref{sec3}. Our case study using both methods on the hot Jupiter HD 189733 b is presented in Section~\ref{sec4}, which shows significant signs of CO, H$_2$O and HCN. We subsequently discuss their sensitivity to detrending methods and hyperparameters. We briefly summarize our results in Section~\ref{sec5}, and discuss approaches to improve the robustness of detrending methods and confidence in detections of molecular species.

\section{Data Reduction and Analysis}\label{sec2}

We use our custom-built pipeline (X-COR 2.0) for analysis of phase-resolved high-resolution spectroscopy of exoplanetary atmospheres. The pipeline involves modular elements for detrending, cross-correlation, and significance  metrics for chemical detections. A previous version of the pipeline was used in the recent high-resolution study of the hot Jupiter HD 209458 b \citep{hawker2018}. In the present work, we explore different detrending methods on specific datasets and investigate the key dependencies in each of them for making molecular detections. The steps common to all the methods are outlined in this section. The different detrending methods and results are discussed in Sec. \ref{sec3} and \ref{sec4}. 

Our analysis consists of four stages. First, we reduce nodding spectra using instrument-specific reduction tools. Second, we generate atmospheric models, from which we obtain spectral templates. Third, we detrend the data to remove telluric contamination and the stellar signal. Finally, we cross-correlate the model template with the detrended data to identify the planet signal. Here we describe each of these stages, with the HD 189733 system as a test case.

\subsection{Data and Initial Reduction}

We use data from observations of HD~189733 taken by the Cryogenic High Resolution Infrared Echelle Spectrograph \citep{kaeufl2004} (CRIRES) mounted at Nasmyth focus A of the Very Large Telescope UT1, located in Cerro Paranal, Chile as part of the CRIRES survey of hot Jupiter atmospheres \citep{2011IAUS..276..208S}. We obtain raw two-dimensional nodding exposures of HD~189733 and their associated calibration files from the ESO Science Archive. The observations were taken for spectral ranges $ 3.1805 < \lambda/\mu$m $< 3.2659 $ and $ 2.2875 < \lambda/\mu$m $< 2.3452$, on the nights of 1 August 2011 and 13 July 2011 respectively. CRIRES has four Aladdin III detectors each with 1024$\times$512 pixels and separation between adjacent chips of $\sim$250 pixel gaps. There are 96 spectra in the 3.2$\mu$m band and 220 spectra in the 2.3$\mu$m band, consisting of 5$\times$30s and 4$\times$15s exposures. Both datasets span orbital phases of $0.38 \lesssim \phi \lesssim 0.48$ over 5 hours of observing time. The telescope was nodded along the 0.2" slit in a standard ABBA pattern, such that pairs of spectra may be combined in a way that accurately subtracts the background. This process reduces the effective resolution of CRIRES to $\sim$87,000.

In order to benchmark the performance of our detrending algorithms, we additionally analyze observations of Tau Boo, presented and discussed by \citet{brogi2012}. The hot Jupiter Tau Boo b orbits a F6V star with magnitude V = 4.5, and has a period of $P = 3.31$ days. Nodding frames of the system were obtained on the nights of April 1, 8 and 14, 2011, covering the wavelength range 2.2875$\mu$m to 2.3454$\mu$m. We choose this dataset based on the very high $6.2\sigma$ detection significance of CO presented in \citet{brogi2012}, and the large quantity of exposures (totaling over 18 hours). The 452 individual reduced spectra cover the planetary orbit of $0.37 \lesssim \phi \lesssim 0.63$.

\begin{figure*}
\centering
\includegraphics[width=\linewidth,trim={10cm 14cm 10cm 14cm},clip]{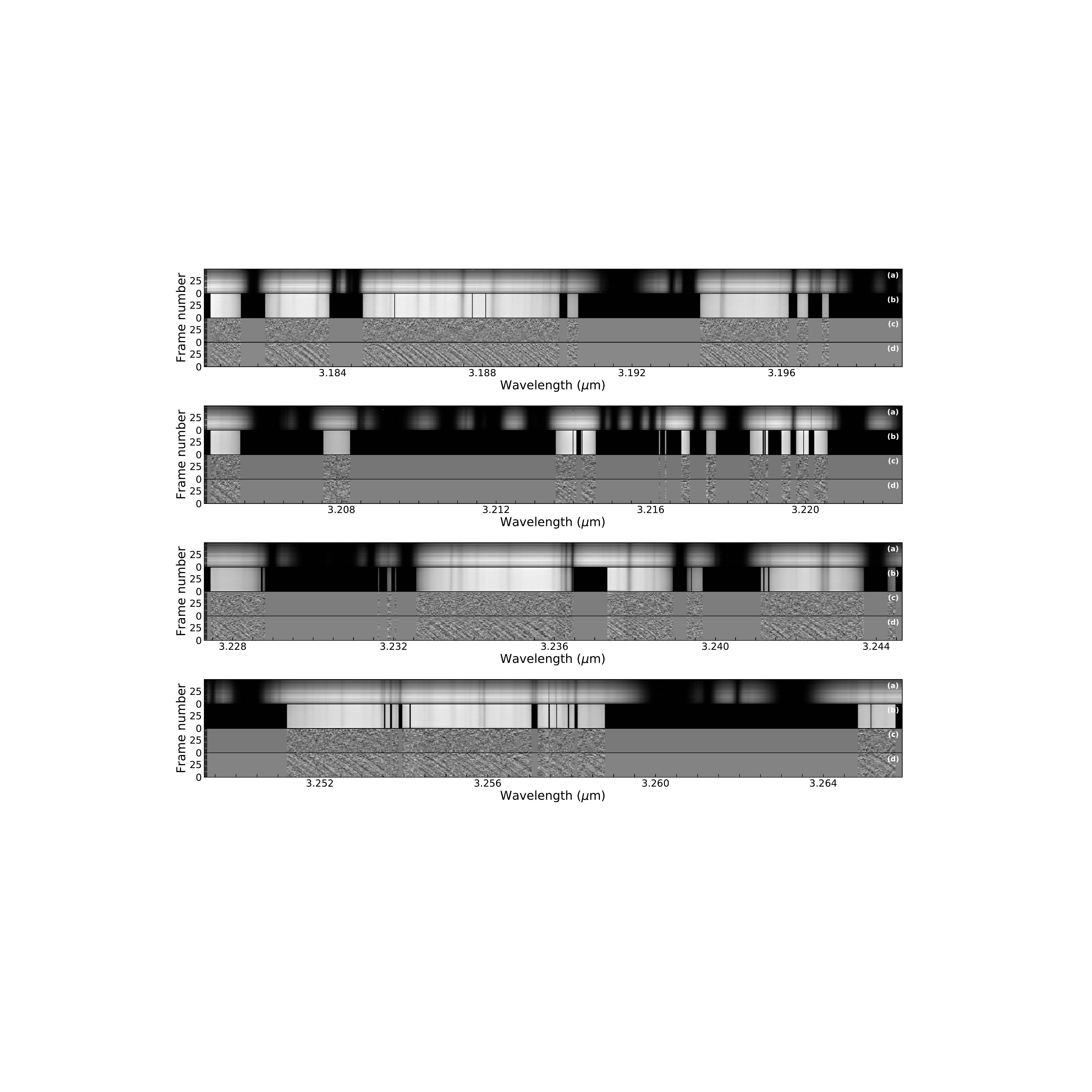}
\caption{The stages of detrending, for detectors 1-4 and the set of observations taken on 1 August 2011. The x-axis corresponds to wavelength, and the y-axis corresponds to frame number, increasing in time. Panel a: spectra immediately after reduction of nodding frames. Heavy telluric contamination is evident (e.g. at 3.1915 $\mu$m). 
Lower instrumental throughput, caused by a combination of poor seeing, inaccurate pointing, or lower sky transparency, manifests as dark horizontal bands. Panel b: reduced spectra after wavelength calibration, alignment, additional cleaning, normalization, and masking. This image (excluding masked regions) is the input of our detrending algorithm. Panel c: data subject to column-wise mean subtraction, the optimal number of SYSREM iterations, a 15-pixel standard-deviation high-pass filter, and column-wise standard-deviation division. Panel d: the same as in Panel c, but with the injection of our planet model at 20x its nominal strength prior to detrending. The preserved planetary absorption features appear as dark trails which stretch over $\sim0.0008 \: \mu$m.}
\label{fig:detrendhcn_plot}
\end{figure*}

\begin{figure*}
\centering
\includegraphics[width=\linewidth,trim={10cm 14cm 10cm 14cm},clip]{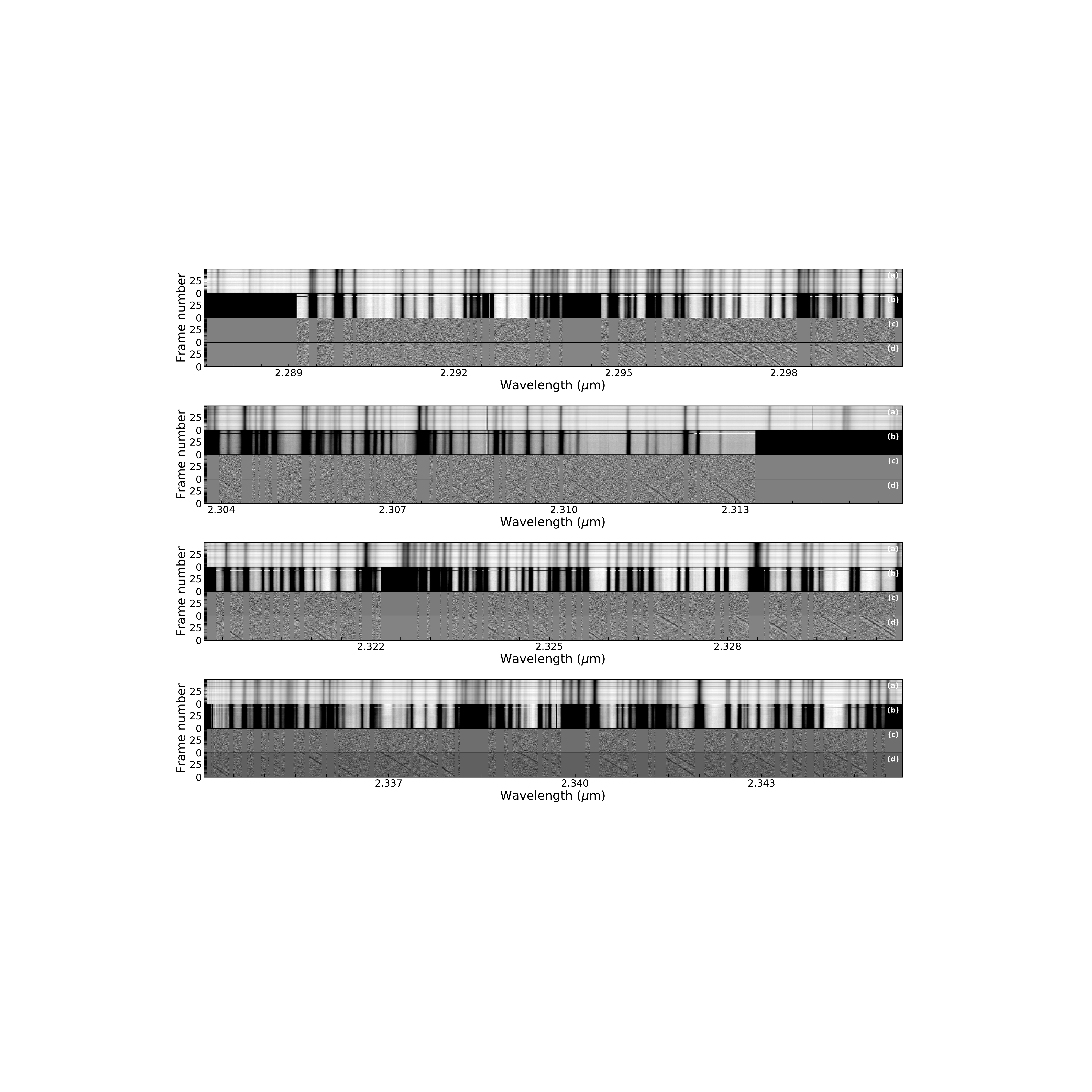}
\caption{The stages of detrending, for detectors 1-4 and the set of observations taken on 13 July 2011. The x-axis corresponds to wavelength, and the y-axis corresponds to frame number, increasing in time. Panel a: spectra immediately after reduction of nodding frames. Heavy telluric contamination is evident (e.g. at 2.3285 $\mu$m). 
Lower instrumental throughput, caused by a combination of poor seeing, inaccurate pointing, or lower sky transparency, manifests as dark horizontal bands. Panel b: reduced spectra after wavelength calibration, alignment, additional cleaning, normalization, and masking. This image (excluding masked regions) is the input of our detrending algorithm. Panel c: data subject to column-wise mean subtraction, the optimal number of SYSREM iterations, a 15-pixel standard-deviation high-pass filter, and column-wise standard-deviation division. Panel d: the same as in Panel c, but with the injection of our planet model at 20x its nominal strength prior to detrending. The preserved planetary absorption features appear as dark trails which stretch over $\sim0.0007 \: \mu$m.}
\label{fig:detrendco_plot}
\end{figure*}

For each spectral range, we treat data from the four CRIRES detectors separately throughout the data reduction and analysis. We take a similar approach to other studies of hot Jupiter atmospheres at high spectral resolution \citep{birkby2013,birkby2017,nugroho2017} using methods common in the literature for data reduction and analysis. The \texttt{Esorex} pipeline (from the ESO CRIRES reduction kit v2.3.4) is used to perform flat-fielding and corrections for gain non-linearities and bad pixels. In addition, the pipeline combines pairs of nodding frames and performs optimal extraction \citep{horne1986} of one dimensional spectra. For subsequent processing, we use custom-built routines written in Python 2.7. We identify remaining bad pixels, correct them via linear interpolation of neighbours, and align all spectra to the highest signal-to-noise ratio spectrum. The alignment is performed via linear interpolation between data points, and maximizing cross-correlation with the reference spectrum on a grid of resolution one-tenth of a pixel. We match telluric absorption lines with features in an ATRAN \citep{lord1992} synthetic transmission spectrum and fit a 3$^{\rm rd}$-order polynomial to calibrate the wavelength grid. The calibration is performed on a sub-pixel grid. The final uncertainty in the resulting wavelength solution is estimated to be $\sim$0.5 km s$^{-1}$ from the centroids of the feature matching with ATRAN. We normalize the baseline level of each spectrum to account for variations in seeing throughout the night. This variation is modelled as the average variation of bright columns away from telluric features and removed by division. The result is a two-dimensional matrix with rows corresponding to individual, time-ordered spectra; the x-axis represents wavelength, and the y-axis represents time (or orbital phase). The top panels of Figures \ref{fig:detrendhcn_plot} and \ref{fig:detrendco_plot} depict this array for 3.2$\mu$m and 2.3$\mu$m observations of HD 189733 b respectively.

\subsection{Modelling}

The atmospheric models are generated using the GENESIS code \citep{gandhi_2017}. The emergent spectrum of the planet is computed using the Feautrier method for radiative transfer and line-by-line opacity calculations. The molecular cross sections for each of the species are calculated using the latest line lists discussed below \citep{rothman_2010, tennyson_2016}. The spectra are generated at a resolution of R$\gtrsim$300,000 in both of the CRIRES wavelength ranges. The details of the model are described below. First we describe the radiative transfer theory used in GENESIS, followed by the molecular line list calculations, and the explored P-T profiles and abundances. 

The specific intensity of radiation $I$ travelling at an angle $\theta$ relative to the vertical is given by

\begin{equation}
\mu \frac{\partial I_{\mu,\nu}}{\partial \tau_{\nu}} = I_{\mu,\nu} - S_{\nu},
\end{equation}

where $\tau_\nu$ is the frequency dependent optical depth, $\mu = \cos\theta$ and $\nu$ denotes the frequency. Here, $S_\nu$ refers to the angle-independent source function of radiation. By defining $j_{\mu,\nu} \equiv \frac{1}{2}(I(\mu) + I(-\mu))$, this equation can be recast into the form

\begin{equation}
\mu^2 \frac{\partial^2j_{\mu,\nu}}{\partial \tau^2} = j_{\mu,\nu} - S_\nu.
\end{equation}

This recast form offers second order accuracy in the number of atmospheric layers and also reduces the number of angle points $\mu$ required for an accurate solution \citep{gandhi_2017, hubeny_2017}. This second order equation is solved for each layer of the atmosphere to determine the radiation field. The emergent flux from the top of the planet is then given by

\begin{equation}
F_{p,\nu} = 2 \pi \int_{0}^{1} \mu I_\mathrm{top,\nu}(\mu) d\mu.
\end{equation}

We adopt 6 angles and use Gaussian quadrature to perform the integral. The flux received from the planet at Earth is then given by $F_{p,\mathrm{Earth}} = F_p \frac{R_p^2}{d^2}$ with $R_p$ the radius of the planet $d$ the distance to the system.

The model atmosphere is divided into 150 layers evenly spaced in log(P) between $100-10^{-8}$ bar in hydrostatic equilibrium and local thermodynamic equilibrium. Given the H$_2$/He dominated atmosphere, the mean molecular mass is taken to be 2.24. The opacity in the atmosphere is contributed by the molecules considered (CO, H$_2$O and HCN) along with continuum opacity due to collision induced absorption (CIA) from molecular hydrogen and helium. We assume a wavelength spacing of 0.01cm$^{-1}$ and compute spectra between 2.25-2.37$\mu$m and 3.15-3.30$\mu$m for both spectral ranges.

We generate high resolution cross sections for each of the species in our analyses using the latest line lists available. The H$_2$O and CO line lists are obtained from the HITEMP database \citep{rothman_2010}, and for HCN we adopt the EXOMOL high temperature line list \citep{harris_2006, barber_2014, tennyson_2016}. The molecular cross sections for each species are calculated from the line lists at 0.01cm$^{-1}$ wavenumber spacing over both spectral ranges using the methods of \citet{gandhi_2017}. This corresponds to a spectral resolution of R$>$300,000 for the 3.18-3.27$\mu$m range and R$>$400,000 in the 2.28-2.35$\mu$m range. We choose a grid over 8 pressures ranging from $100-10^{-5}$ bar evenly spaced in log(P), and 16 temperatures ranging from 300-3500K \citep{gandhi_2017}. From these cross sections we interpolate to the pressure and temperature in each layer of the model atmosphere for calculation of the emergent spectrum. As well as molecular opacity we also include collisionally induced absorption (CIA) from H$_2$-H$_2$ and H$_2$-He interactions \citep{richard_2012}.

\begin{figure}
	\includegraphics[width=\columnwidth]{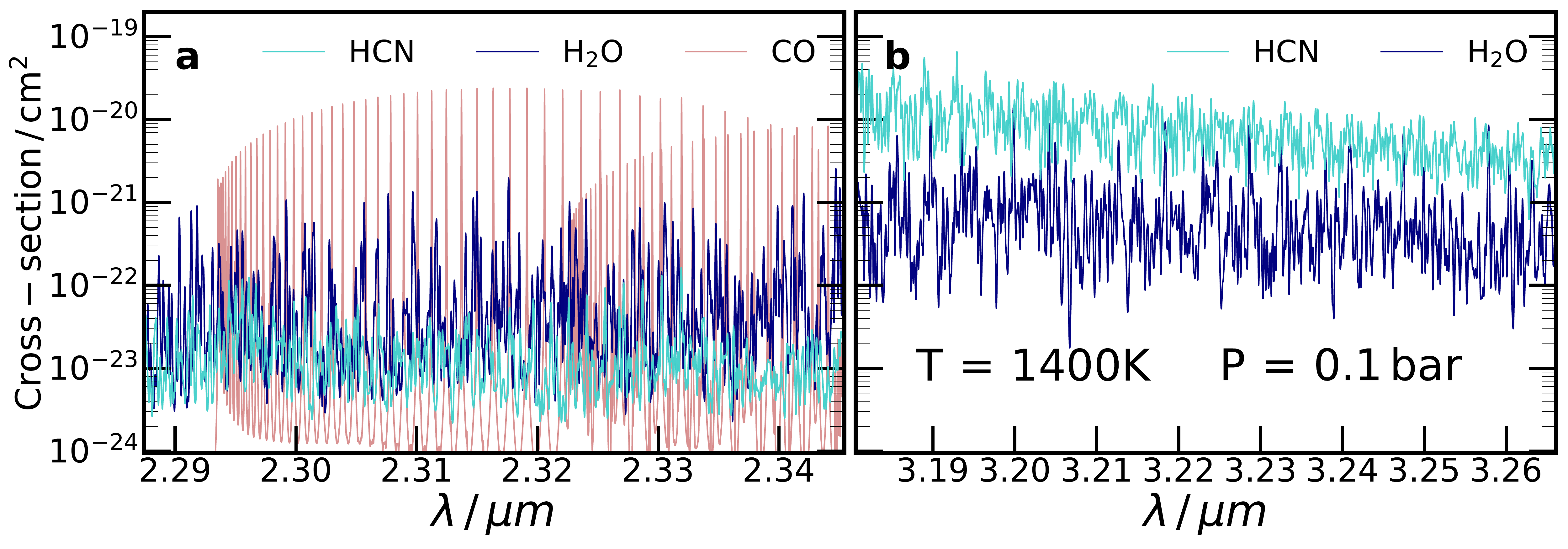}
    \caption{Molecular cross sections for each species. These are generated in the 2.28-2.35$\mu$m (panel a) and 3.18-3.27$\mu$m (panel b) CRIRES spectral ranges and smoothed for clarity in the figure. The cross sections are shown for HCN (turquoise), H$_2$O (blue) and CO (red).}
    \label{fig:cs}
\end{figure}

Figure \ref{fig:cs} shows the molecular cross sections of H$_2$O, CO and HCN over the two spectral ranges. These are shown at a representative photospheric pressure of 0.1 bar and a temperature of 1400K, corresponding to the best-fit model. In the 2.28-2.35$\mu$m range, CO has the dominant molecular cross section due to its strong absorption lines. However, in the 3.18-3.27$\mu$m range, the HCN and H$_2$O cross sections dominate over CO (not shown in Figure \ref{fig:cs} b due to its negligible opacity in the 3.2$\mu$m waveband).

\begin{figure}
	\includegraphics[width=\columnwidth,trim=2cm 0 0cm 0,clip]{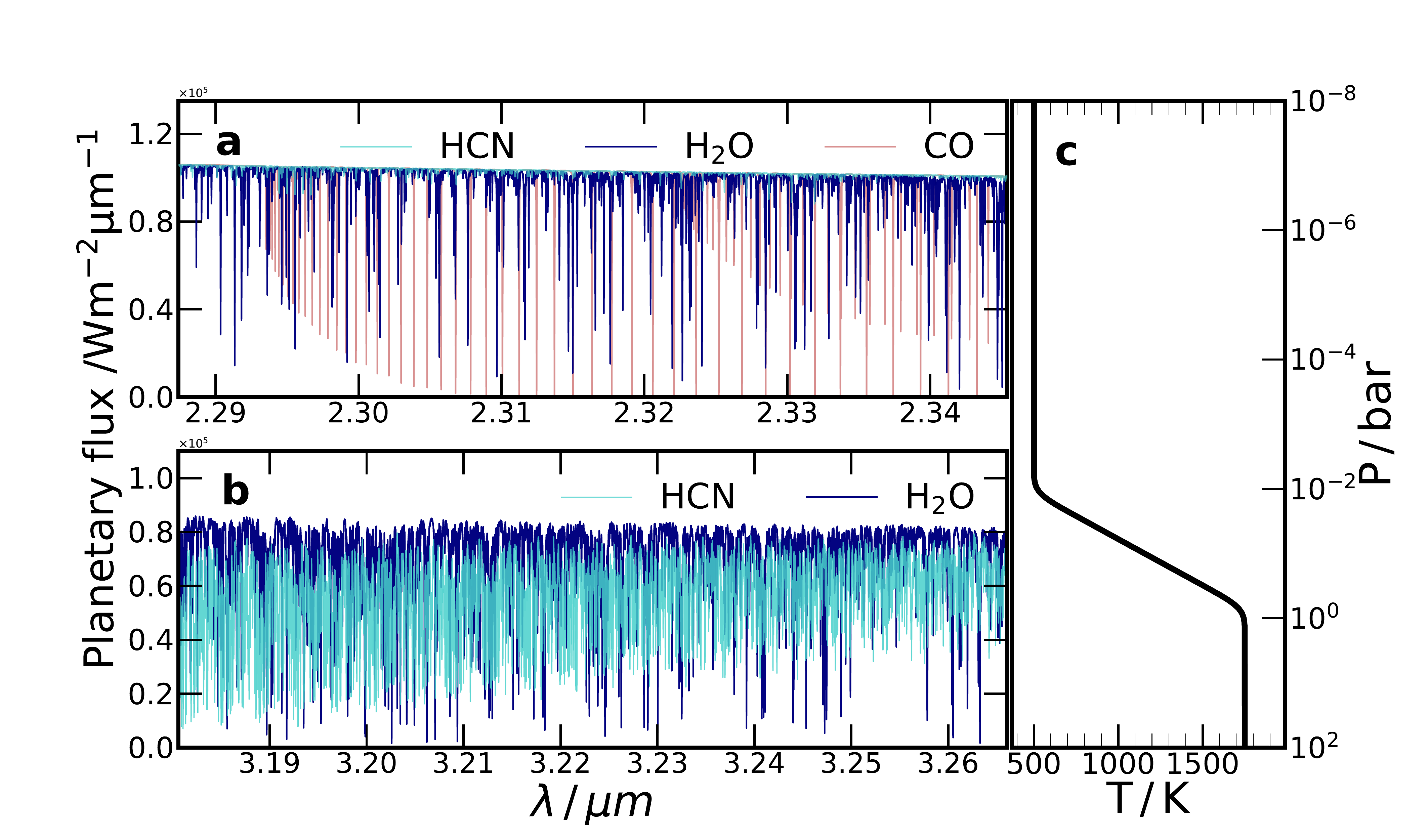}
    \caption{Planetary spectra of the model atmosphere of HD~189733 b with different molecular species present. Panel a shows the 2.28-2.35$\mu$m spectral range and panel b shows the 3.18-3.27$\mu$m range. Spectra were generated with HCN (turquoise), H$_2$O (blue) and CO (red) with a wavenumber spacing of 0.01cm$^{-1}$ using the GENESIS code \citep{gandhi_2017}. Panel c shows the P-T profile used to generate the spectra.}
    \label{fig:flux}
\end{figure}

Model spectra corresponding to HCN and H$_2$O in the two spectral ranges are shown in Figure \ref{fig:flux}, along with a pressure-temperature (P-T) profile used in our analysis. The HCN spectrum has significant features in the 3.18-3.27$\mu$m region due to its strong molecular cross-section. However, all of the spectra show absorption features from the strongest molecular lines of each molecule. These absorption features occur due to the P-T profile in which the temperature is decreasing monotonically upward. The distinct line positions and strengths for each species are used to generate templates for the cross-correlation analysis.

\subsection{Detrending}

We briefly discuss the detrending step below, and perform a thorough investigation of different detrending methods in Section~\ref{sec3}. The reduced spectra are dominated by the stellar signal and deep telluric absorption lines caused by Earth's atmosphere. However, these features remain nearly constant with time throughout the night of observation, while the planet signal incurs Doppler shifts from its changing radial velocity (RV). As such, robust detrending methods are required to remove the non-planet signals from the data. Two key approaches have been used in previous studies for detrending \citep{birkby2018}. We implement algorithms based on modelling airmass and telluric residuals, as well as the unsupervised SYSREM \citep{tamuz_2005} algorithm. A single iteration of SYSREM identifies and subtracts time-invariant signals, systematics and environmental effects. After a sufficient number of iterations, only the time-varying planet signal remains, along with Gaussian noise. This process is illustrated in the Figures~\ref{fig:detrendhcn_plot} \& \ref{fig:detrendco_plot}. We compare the performance of airmass-based detrending and SYSREM in Section~\ref{sec4}. Detrending typically involves the additional application of a high-pass filter, and normalizing the residuals by their uncertainty.

\subsection{Cross Correlation Analysis}

We obtain cross-correlation templates from model spectra of HD~189733 b atmosphere by fitting the amplitudes and positions of the strongest spectral lines with narrow Gaussian profiles. We did this for models with CO, H$_{2}$O, HCN, and combined HCN $+$ H$_{2}$O. Previous studies \citep{snellen2010, brogi2012} motivate this technique since it ignores the continuum baseline, noisier weak lines, and broad-band components (removed earlier by high-pass filter) thus improving the correlation with intrinsic spectral features. 

We cross-correlate each detrended spectrum with the spectral templates. The templates are Doppler shifted from -250 km s$^{-1}$ to 250 km s$^{-1}$ in steps of 1 km s$^{-1}$ to probe a large grid of potential planet RVs. The result is a Cross-Correlation Function (CCF) matrix. We then sum the CCFs from all four detectors. 

\begin{figure}
\centering
\includegraphics[width=\linewidth,trim={0 0 0 3cm},clip]{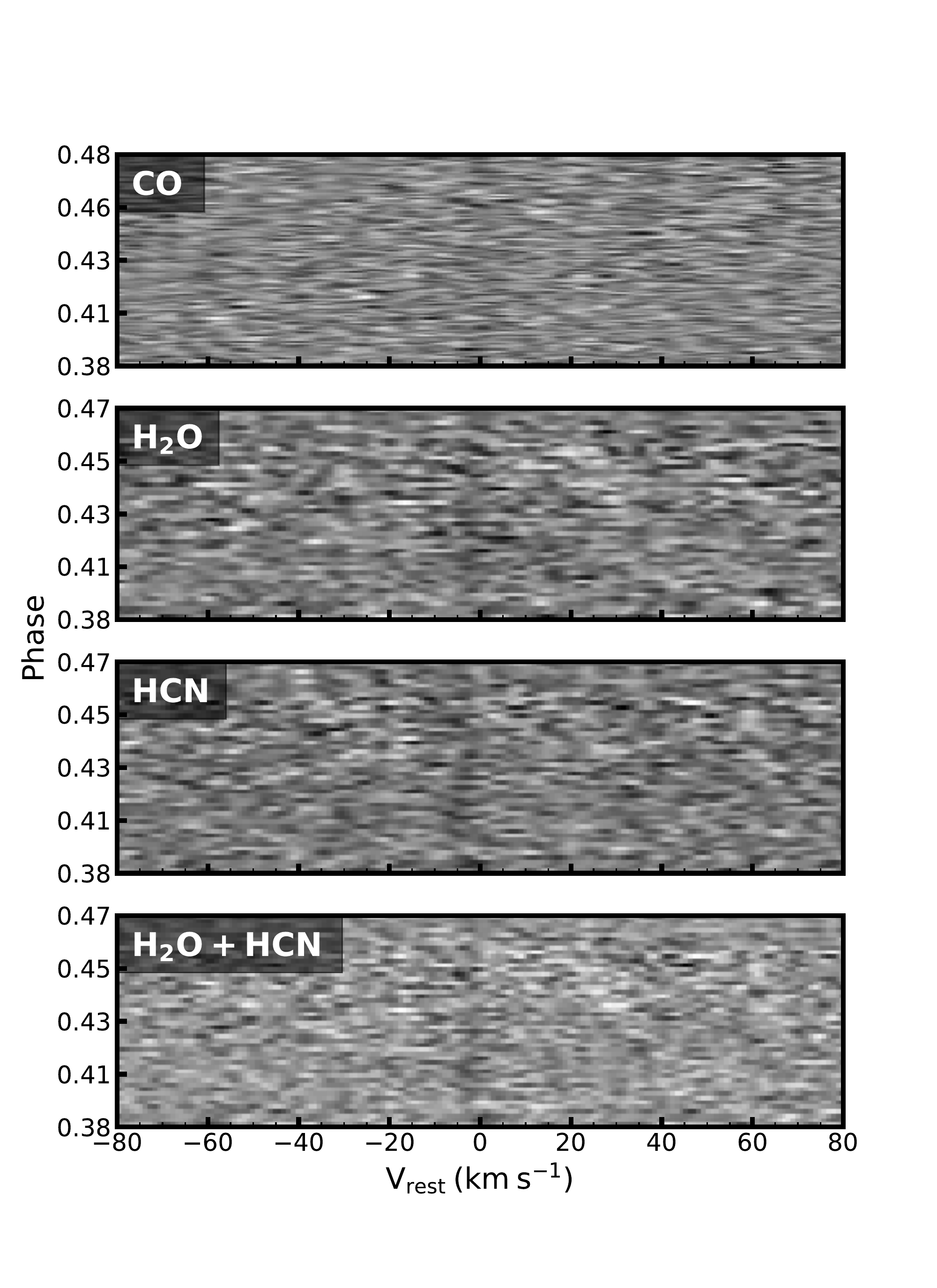}
\caption{Cross-correlation values as a function of orbital phase and velocity for HD 189733 b. The CCFs are shifted into the planetary rest frame by the $K_p$ of the highest detection significance, for each model template (CO , H$_2$O, HCN, and the combination H$_2$O $+$ HCN). Darker pixels correspond to higher values. Each CCF contains a dark vertical trail at approximately the known systemic velocity of 2.3 km s$^{-1}$ from strong cross-correlation between the model template and intrinsic features in the data. The data are obtained in the 2.3$\mu$m band for CO and in the 3.2$\mu$m band for HCN and H$_2$O.}
\label{fig:ccf_plot}
\end{figure}

We refrain from making {\it a priori} assumptions regarding the value of the planetary orbital semi-amplitude, $K_p$. Rather, we sample $K_p$ values ranging from 20 km s$^{-1}$ to 180 km s$^{-1}$ in steps of 1 km s$^{-1}$. For each $K_p$, we shift the CCF by $K_p\sin\phi$ with the phase determined by the orbital period $P = 2.2186$ days along with a reference crossing time $t_{\phi=0} = 53628.889$ MJD \citep{triaud2009}. Under the correct transformation, the CCF is shifted into the planetary rest frame. The column corresponding to $V_{sys}$ contains pixels where the template has strong cross-correlation with data. We then sum all cross-correlation values within a 3-column wide sliding window. The results for all values of $K_p$ are then presented in an array, normalised by its standard deviation to give a signal-to-noise (SNR) array. A high SNR signal at the known $V_{sys}$ and planetary $K_p$ thus indicates the presence of a given molecular species. Figure~\ref{fig:ccf_plot} depicts the CCFs, transformed according to their corresponding peak $K_p$ values, following the SYSREM detrending method. Dark trails are visible at the known value of $V_{sys}$. We present detection significances for each template search and both detrending methods in the following section. The SNR significances are computed over the grid of $K_p$ and $V_{sys}$.

\section{Detrending Methods}\label{sec3}

In this section, we investigate two different detrending methods used for removing non-planetary components in the data. Robust detrending methods are central to chemical detections of exoplanetary atmospheres using high-resolution spectroscopy. In most high-resolution studies, the expected planet-to-star flux ratio $f_p/f_s$ is typically between $10^{-4}$ and $10^{-3}$. The signal-to-noise ratio of the data is typically on the order of 10-100. Thus the planet signal is well below the Poisson uncertainty of the stellar flux, and is further obscured by telluric contamination from Earth's atmosphere, environmental factors such as airmass and temperature, and systematics affecting Adaptive Optics and CCD performance. Previous detections exploit the approximate time- or wavelength-invariance of such nuisance signals. The planet signal however is Doppler shifted by tens of km/s, corresponding to tens of pixels on the CRIRES CCD. By `detrending' the dataset, one removes all components except the Doppler shifted planet signal. 

There are various detrending methods currently used in the literature, primarily focused on the removal of telluric contamination and airmass variation. The two most common methods are: (1) removing a polynomial fit to the time axis (columns of the data) which captures airmass variations, and repeating for remaining residuals in H$_2$O and CH$_4$ telluric lines \citep{brogi2012,brogi2013,brogi2014,brogi2016,schwarz2015}; and (2) subtracting a low-rank approximation of the data based on principal component analysis with SYSREM or Singular Value Decomposition (SVD) \citep{birkby2013,dekok2013,birkby2017,nugroho2017}. Another less-common method not investigated in this work involves directly fitting telluric zones with tools such as ESO MOLECFIT software \citep{smette2015,rodler2012,lockwood2014}. Some previous studies apply a high-pass filter to each resultant spectrum in order to remove low-order variation \citep{brogi2013,birkby2017}. Noise and residuals remain after detrending. It is common to then use a cross-correlation approach to identify the planet signal as discussed in the previous section. While some studies mask only a few damaged columns of data from the analysis \citep{brogi2012,schwarz2015}, other studies employ detrending methods which rely on the masking of the worst telluric regions initially \citep{birkby2013,birkby2017}. There is little work in the literature investigating the importance of this step. The masking can be done quantitatively by considering low flux and high variance columns or more qualitatively by eye. Here we apply different detrending methods to the dataset and investigate different masking criteria.

\subsection{Fitting a function of Airmass}

Airmass variation affects the overall extinction caused by the atmosphere, and the depth of methane and water telluric absorption lines. Early high-resolution detections of CO in HD 209458 b \citep{snellen2010} and Tau Boo b \citep{brogi2012} involved fitting and removing trends due to airmass. This step prevents atmospheric affects from dominating the cross-correlation function. We investigate the effectiveness of this method and its variations on the 2.3$\mu$m observations of Tau Boo b \citep{brogi2012}. We use the following detrending prescription: we mask a small number of columns that are affected by systematics or suffer significant telluric contamination. For each remaining column, we determine a linear fit between the column fluxes and the geometric airmasses of the observations, and subsequently divide the column by the fit. Strong residuals remain in columns corresponding to deep telluric absorption. We sample several of these columns. Then, we fit a linear combination of these samples to the fluxes in each column of data, and divide by the fit. We next apply a high-pass Gaussian filter of standard deviation 15 pixels to each row of the data in order to remove low-order variation. Finally, we normalize each pixel by its uncertainty (the sum of extraction error and Poisson uncertainty in quadrature) to prevent noisy pixels from affecting the cross-correlation function. 

The initial masking step prevents bad columns from influencing the detrending process and the cross-correlation, and is common practice in high-resolution literature \citep{brogi2012,birkby2013,birkby2017}; however there is no consistent criteria for classifying columns as bad. Before detrending a given detector, we mask columns based on two parameters, $p_v$ and $p_m$. We mask columns with variance in the top $p_v$ percentile of all column variances. Columns with exceptionally high variance are expected to result from systematics or artifacts. We also mask columns with mean in the bottom $p_m$ percentile of all column means. These are the columns which are most affected by tellurics such that little flux is transmitted. We also mask $1\%$ of data from each end of each detector to remove edge effects from wavelength-calibration. Optimal values for $p_v$ and $p_m$ depend on the quality and extent of the data. We explore the sensitivity of detrending to masking parameters in Section~\ref{sec4}.

\begin{figure}
	\includegraphics[width=\columnwidth,trim={9cm 0 2cm 0},clip]{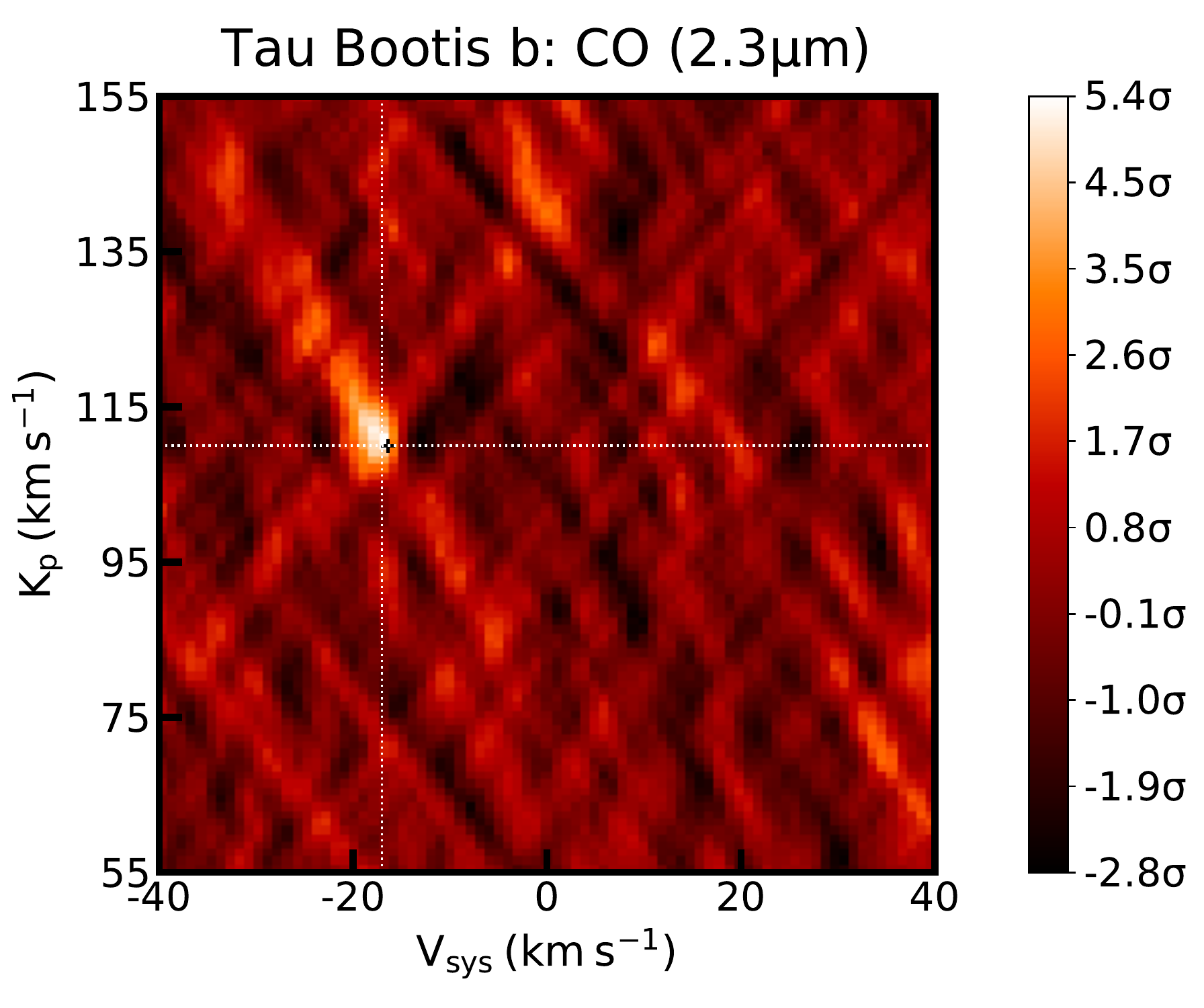}
\caption{Detection of CO in Tau Bootis b using data combined from three nights of observation. The white cross-hairs denote the peak detection significance, while the black plus indicates the expected planetary $K_p$ and $V_{sys}$, consistent with \citet{brogi2012}.
   }
    \label{fig:tb_airmass}
\end{figure}

\begin{figure}
	\includegraphics[width=\columnwidth,trim={0 0 0 0},clip]{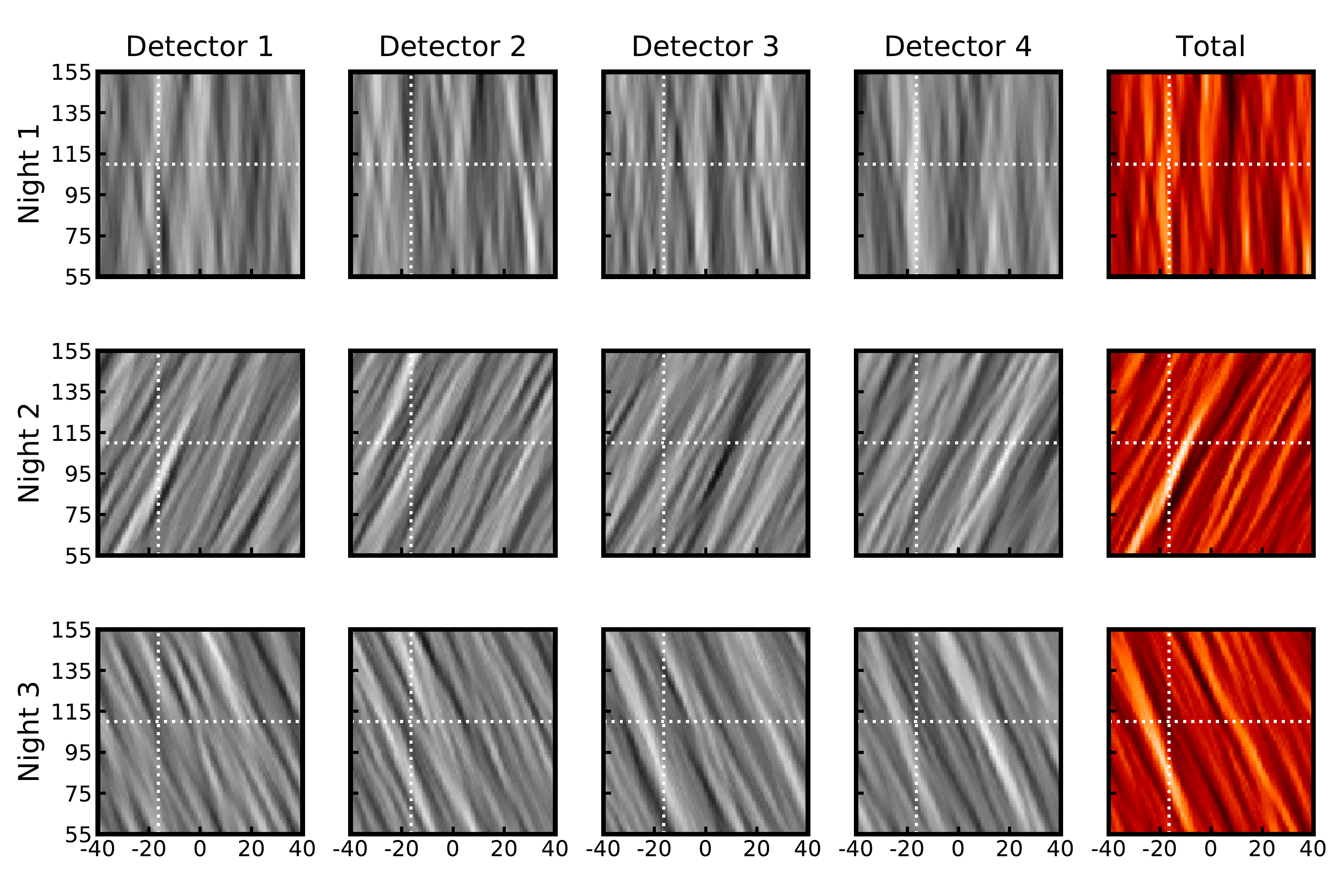}
    \caption{
    Detection significances of CO for each night of observation of Tau Boo. The gray panels show contributions from individual detectors. The red panels show the detection significances combining all data for each night. The x-axis of each plot corresponds to $V_{sys}$ in km s$^{-1}$, and the y-axis corresponds to $K_{p}$ in km s$^{-1}$. The color scale for the grid of plots ranges from values of $-4\sigma$ (darkest) to $+4\sigma$ (lightest). The white cross-hairs in each plot indicate the expected planet $K_p$ and $V_{sys}$ location.}
    \label{fig:tb_grid}
\end{figure}

We apply the detrending procedure to the 2.3$\mu$m observations of Tau Bootis b, and cross-correlate with a CO template consisting of narrow Gaussian profiles at the locations of each absorption feature. We sum the contributions from all four detectors, and scan the $K_p$ and $V_{sys}$ space as described in Section~\ref{sec2}. The peak detection significance varies by about $\pm1.0\sigma$ depending on the selection of residual telluric columns. Fitting and removing too many higher-order residuals degrades the planet signal, while selecting too few allows the residuals to affect the cross-correlation function. We use an automated procedure to robustly identify and remove residuals: the K-Means clustering algorithm from the \texttt{scikit-learn} Python package classifies columns based on their residuals following airmass detrending. We linearly fit the mean of each cluster to each column of the data, and remove it by division. The algorithm acts on half of the columns with the highest variance, so that it only finds clusters associated with residuals. We find 8 clusters yields the highest detection significance at the expected location of the planet signal. Applying this procedure, coupled with a high-pass filter and pixel uncertainty-normalization, we successfully confirm the detection of CO presented in \citet{brogi2012} at a significance of $5.4\sigma$ (Figure~\ref{fig:tb_airmass}). The detection peak at $K_p = 110$ km s$^{-1}$ and $V_{sys} = -17$ km s$^{-1}$ directly corresponds with the expected planetary orbital parameters \citep{brogi2012}, following the application of a $\Delta\phi = +0.0068$ phase shift. This shift is within the uncertainty of the orbital period and reference time, and precisely agrees with the shift applied by \citep{brogi2012}. A single set of observations produces elliptical contours due to the degeneracy between $K_p$ and $V_{sys}$ (Figure~\ref{fig:tb_grid}). Combining the multiple nights of observation yields a criss-cross pattern which tightly constrains the contour of highest detection significance. We also cross-correlate with an H$_2$O template, but do not find significant evidence of water absorption in this dataset.

\subsection{PCA based methods and SYSREM}

High-resolution spectroscopy can be extended to searches for more complex molecules. The 3.2$\mu$m CRIRES wavelength band is a natural area to search for H$_2$O given its high opacity in this band \citep{gandhi_2017}. However, two difficulties arise: water spectral features, while many in number, are significantly weaker than those from CO in this band. Additionally, the 3.2$\mu$m wavelength regime is heavily affected by telluric contamination. This prompts the use of alternative detrending methods.

In a study of HD 189733 b, \citet{dekok2013} detected CO in emission using an approach based on Singular-Value Decomposition (SVD). The SVD algorithm determines singular vectors of the spectral array, and their corresponding singular values. The spectral array may be approximated as a linear combination of the top $r$ singular vectors. By subtracting this approximation from the original array, one captures and removes information in the array that is well represented by a vector (i.e. effects which are constant in the time or wavelength direction). This method is particularly useful for modeling and removing higher-order trends from the data. In another study of HD 189733 b, \citet{birkby2017} detected H$_2$O in emission using SYSREM \citep{tamuz_2005}, a variant of SVD. SYSREM iteratively minimizes the least-squares difference between the original spectral array and its low-rank approximation. SYSREM weights the contributions of individual pixels by their uncertainty (the quadrature sum of optimal extraction error and shot-noise). In the equal-error case, the SYSREM algorithm reduces to principal component subtraction. In this section, we use SYSREM to detrend 3.2$\mu$m observations of HD 189733 b, and perform cross-correlation with an H$_2$O molecular template.

\begin{figure}
\centering
\includegraphics[width=\linewidth,trim={0 0 0 0},clip]{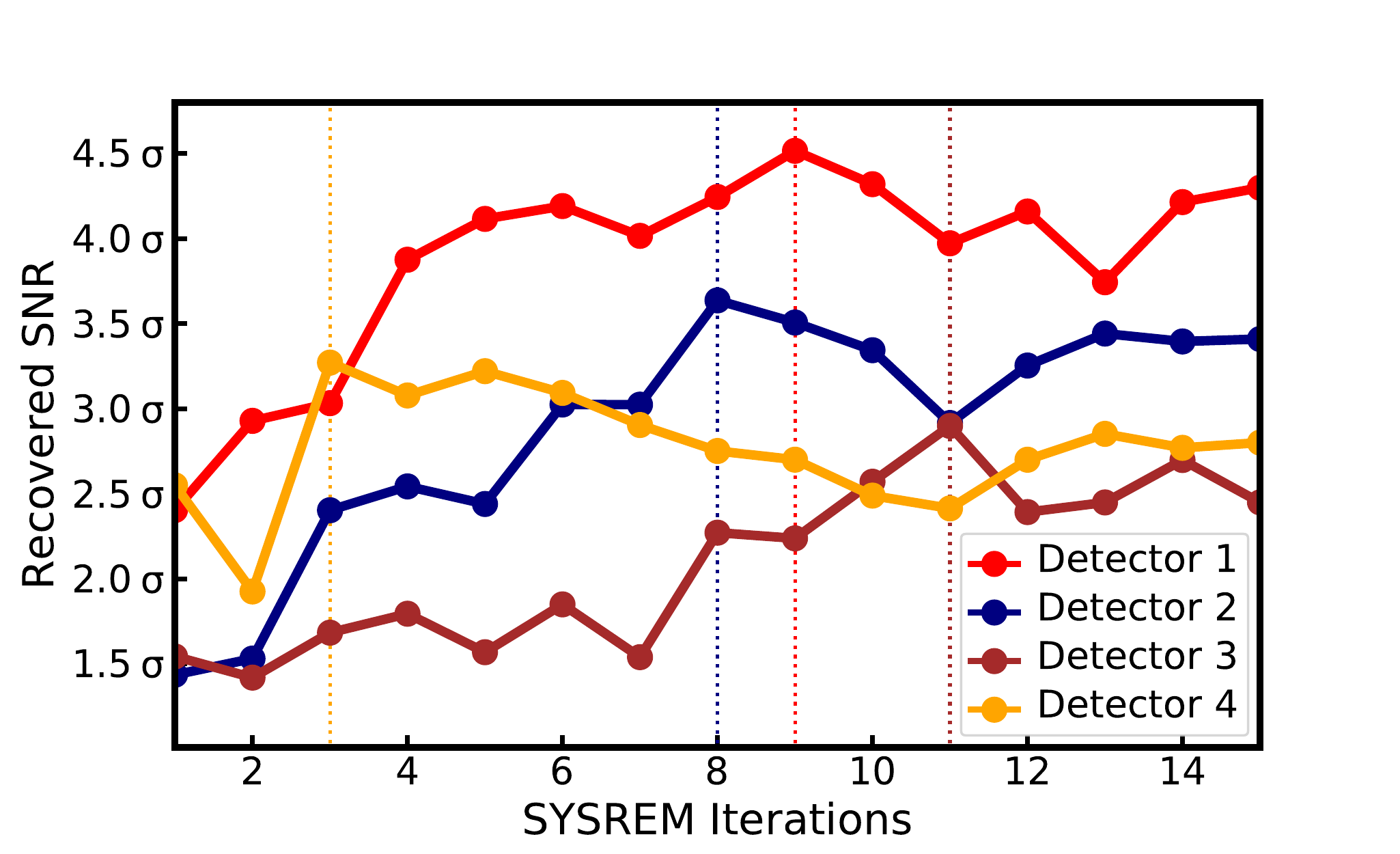}
\caption{Recovered detection SNR for each detector as a function of applied SYSREM iterations. The recovery is based on injection of a model spectrum of HD 189733 b at $1\times$ nominal strength. The vertical dotted lines mark the peak recovery significances for each detector in the 3.2$\mu$m band and the number of SYSREM iterations used in the analysis.
}
\label{fig:sysdet}
\end{figure}

\begin{figure}
\centering
\includegraphics[width=\linewidth,trim={0 0 0 0},clip]{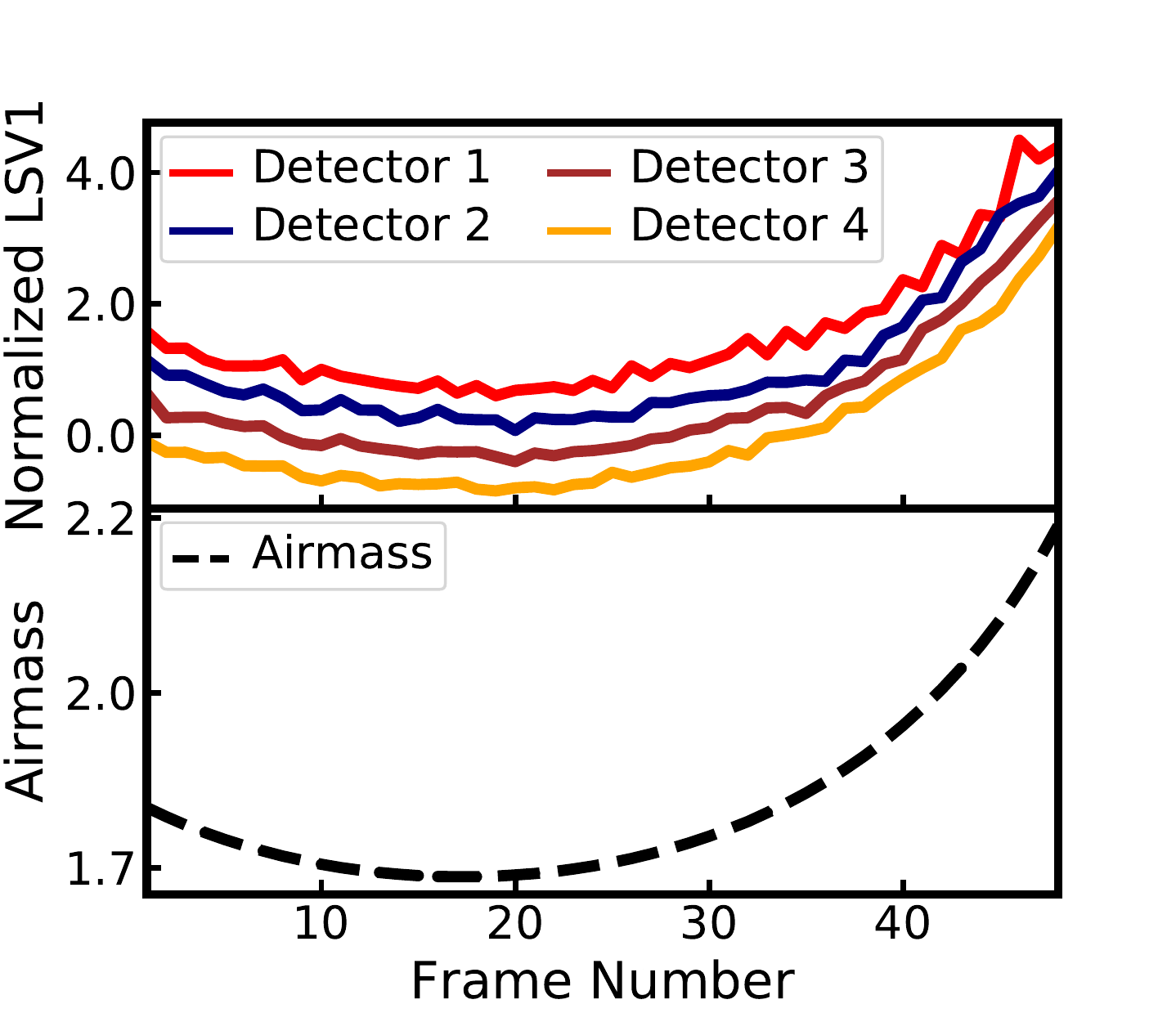}
\caption{The first trend removed by SYSREM. Top: the first left singular vector (LSV) removed by SYSREM in the 3.2$\mu$m dataset, for each detector. The curves for each detector have been normalized and centered, and subsequently offset for clarity. The first SYSREM iteration subtracts a linear scaling of the vector from each column of the detector. Bottom: recorded airmass for each frame. There is a clear correspondence between the first LSV of each detector and the airmass trend.
}
\label{fig:syscomp}
\end{figure}

We mean-subtract all columns to remove the zeroth order principal component, and run SYSREM for N iterations. We choose the number of iterations, as well as the amount of masking, by injecting a planet model at the expected $V_{sys}$ and $K_p$ at 1$\times$ its nominal strength, and maximising the detection significance in recovering the model with the cross-correlation methods described in Section~\ref{sec2}. The optimization is performed for each detector, over a grid of SYSREM iterations ranging from 1-15 and masking parameters $p_v$ and $p_m$. We do this in accordance with previous studies \citep{birkby2017,nugroho2017} because excess SYSREM iterations will eventually remove the planetary signal. We optimize using the HCN molecular template, which allows us to simultaneously obtain high-significance detections of H$_2$O and HCN under the same parameters, as discussed in the following section. We find 9, 8, 11, and 3 iterations are optimal for detectors 1-4 respectively in the $3.2\mu$m band (Figure~\ref{fig:sysdet}). The number of iterations for each detector depends on several factors, such as the number of planetary absorption lines that fall within its coverage and the degree of telluric contamination. The odd-even effect, caused by gain variations between adjacent columns in detectors 1 and 4, may also influence the optimal number of iterations. The first component removed by SYSREM for each detector is shown in Figure~\ref{fig:syscomp}, showing strong correlation with airmass. After using SYSREM, remaining broadband variations are removed by applying a high-pass Gaussian filter of standard-deviation 15.0 pixels to each row. Finally, we divide each column by its standard-deviation, to prevent particularly noisy pixels from dominating the Cross-Correlation Function. We mask between $40-70\%$ of columns for each detector, in addition to 1$\%$ from either end to eliminate edge-artifacts. The detrending steps described above are nearly identical to those presented by \citet{birkby2013},
with the additional optimization of masking levels. The full process of detrending is shown in Figure~\ref{fig:detrendhcn_plot}.

\begin{figure}
\centering
\includegraphics[width=\linewidth,trim={4.5cm 0 1cm 0},clip]{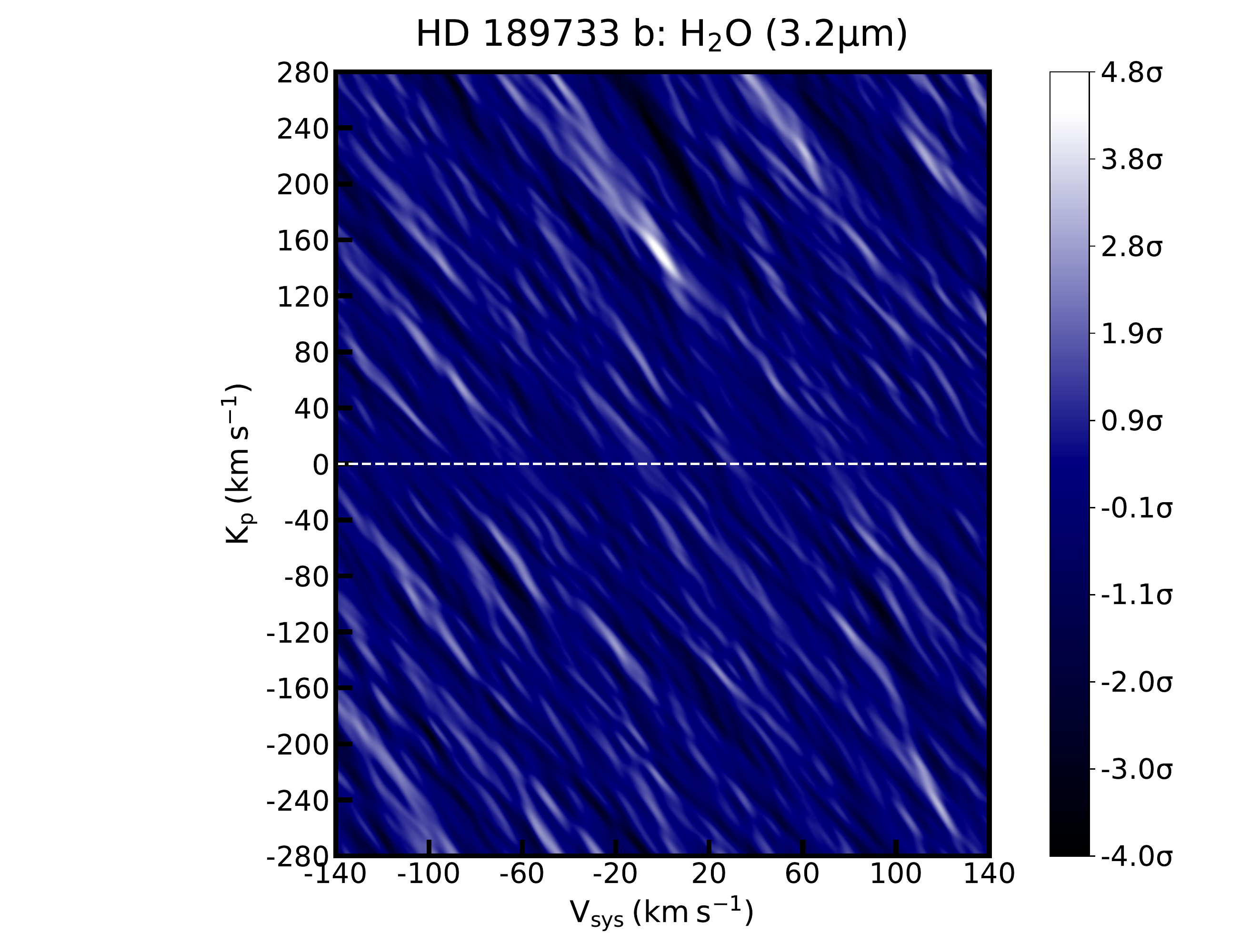}
\caption{Detection significances from cross-correlating the 3.2$\mu$m dataset of HD 189733 b with an H$_2$O model template. There is an evident peak at the planetary $V_{sys}$ and $K_p$, consistent with \citet{birkby2013}. This plot depicts an expanded velocity search range which additionally explores negative values of $K_p$. A dashed white line marks $K_p = 0$.
}
\label{fig:hd_big}
\end{figure}

We detect H$_2$O at 4.8$\sigma$ significance. The SNR peak is at $V_{sys} = -2.0^{+6.0}_{-4.0}$ km s$^{-1}$ and $K_p = 151.0^{+11.0}_{-14.0}$ km s$^{-1}$ (Figure~\ref{fig:hd_big}), agreeing precisely with the expected $V_{sys} = -2.361$ km s$^{-1}$ and $K_p = 152$ km s$^{-1}$ of the planet \citep{triaud2009,bouchy2005}. The detection peak is $\sim 0.8\sigma$ stronger than other spurious correlations and anti-correlations. We perform cross-correlation across a large range of $K_p$ and $V_{sys}$, and expand the $K_p$ search range to negative values. Noise properties of the data indicate that if the detection is a false positive, a peak might be observed at the negative $K_p$ \citep{brogi2014}. However, we do not find such a peak, which improves confidence in the detection. Our detection comes at a comparable significance to the $5.1\sigma$ detection reported in \citet{birkby2013}.

\section{HD 189733 \texorpdfstring{\MakeLowercase{b}}{b}: A Case Study}\label{sec4}

In this section we use the hot Jupiter HD 189733 b as a case study to search for molecules using the two detrending methods -- SYSREM and airmass fitting. Our observations are as described in Section~\ref{sec2}, consisting of time-series spectra obtained with CRIRES in the 2.3$\mu$m and 3.2$\mu$m spectral bands \citep{2011IAUS..276..208S}. We perform cross-correlation searches for CO, H$_2$O and HCN, which are expected to be prevalent in hot Jupiter atmospheres \citep{madhu2012}. We use these datasets as a case study to detect molecules using the two detrending methods. We start with results obtained from SYSREM, and subsequently report results from detrending with airmass. We directly compare results, and discuss the effectiveness of the algorithms in the two wavelength regimes. Then, we conduct a systematic exploration of molecular detection significances, and the robustness between the two methods. This includes determining the sensitivity of detections to detrending hyperparmeters, and the frequency of high-amplitude, spurious correlations and anti-correlations.

\subsection{Molecular Detections}

Using SYSREM, we detect CO with a peak detection SNR of $4.7\sigma$ at $V_{sys} = 0.0^{+3.0}_{-3.0}$ km s$^{-1}$ and $K_p = 148.0^{+8.0}_{-11.0}$ km s$^{-1}$. These are consistent with the literature values \citep{bouchy2005} of $V_{sys} = -2.361$ km s$^{-1}$ and $K_p = 152$ km s$^{-1}$ to which we optimized. Uncertainties correspond to points within a $1\sigma$ contour surrounding the peak significance. The peak H$_2$O significance is 4.9$\sigma$ at $V_{sys} = -2.0^{+6.0}_{-4.0}$ km s$^{-1}$ and $K_p = 151.0^{+11.0}_{-14.0}$ km s$^{-1}$. The peak HCN detection significance is 5.0$\sigma$ at $V_{sys} = -4.0^{+4.0}_{-5.0}$ km s$^{-1}$ and $K_p = 155.0^{+13.0}_{-9.0}$ km s$^{-1}$. An HCN mixing ratio of $10^{-6}$ yields the peak detection significance reported above. The shape of the contours in Figure~\ref{fig:hd_comp} is due to the degeneracy in $K_p$ and $V_{sys}$ when the CCF is nearly aligned to the rest frame. The orientation is determined by $\Delta \phi$. We use the same masking levels and number of SYSREM iterations (9, 8, 11, and 3) presented in the previous section for the 3.2$\mu$m band. We also optimize with our planet model in the $2.3\mu$m band, and find 3, 12, 4, and 13 iterations are optimal. The detections of CO and H$_2$O are consistent with previous studies analyzing these data \citep{dekok2013,birkby2013}. Similar evidence of HCN, also using SYSREM, was recently reported for the hot Jupiter HD 209458 b \citep{hawker2018}.

Using airmass-based detrending, we find evidence for CO absorption with a peak detection SNR of $3.7\sigma$ signal at $V_{sys} = 1.0^{+22.0}_{-9.0}$ km s$^{-1}$ and $K_p = 146.0^{+12.0}_{-37.0}$ km s$^{-1}$. The peak H$_2$O significance is 3.8$\sigma$ at $V_{sys} = -1.0^{+6.0}_{-9.0}$ km s$^{-1}$ and $K_p = 151.0^{+18.0}_{-16.0}$ km s$^{-1}$. The peak HCN detection significance is 3.4$\sigma$ at $V_{sys} = -2.0^{+19.0}_{-19.0}$ km s$^{-1}$ and $K_p = 152.0^{+28.0}_{-40.0}$ km s$^{-1}$. While the detection peaks are consistent with the expected $K_p$ and $V_{sys}$, they are less well-constrained as demonstrated by larger error bars. Detrending the 2.3$\mu$m dataset involved sampling 9 residual columns for the higher-order fit. We use 3 residual columns for the 3.2$\mu$m dataset. We find the optimal airmass-detrending approach involves performing a linear fit with airmass, performing another linear fit with sampled residuals, applying a high-pass filter, and normalizing each column by its standard deviation; however we compare variations below.

We conducted additional tests to validate the significance of our results. We compare the distributions of `in-trail' and `out-of-trail' rest-frame CCF values. The `out-of-trail' values are consistent with Gaussian noise out to $\sim 4.0\sigma$ in a quantile-to-quantile comparison. A Welch $T$-test rejects the hypothesis that the two samples are drawn from the same parent distribution at confidences of 6.12$\sigma$, 6.23$\sigma$, 6.70$\sigma$, 6.81$\sigma$ for CO, H$_2$O, HCN, and the combined H$_2$O $+$ HCN model respectively. However, we choose to report the detection significances above from the more conservative signal-to-noise metric, which is the peak CCF sum normalised by the standard deviation across the $K_p$ and $V_{sys}$ search range. We perform an additional test in which we subtract our HCN model at the planet $K_p$ and $V_{sys}$ and re-inject it at a grid of locations of $K_p \in [40, 160]$ km s$^{-1}$ and $V_{sys} \in [-60, 60]$ km s$^{-1}$, and record the peak significance. We repeat this 100 times, and find an average peak significance of $4.31 \pm 0.84 \sigma$, consistent with the observed significance under our chosen metric. Finally, we note that under the same number of SYSREM iterations, we detect HCN and successfully reproduce the detection of H$_2$O at a significance comparable to that reported in literature \citep{birkby2013}; and combining HCN and H$_2$O features in a single model boosts the detection significance by 0.9$\sigma$ to 5.9$\sigma$ as shown in Figure~\ref{fig:hd_comp}. 

\subsection{Comparison of Detrending Methods}

The detections of H$_2$O, CO, and HCN come at high-significance (SNR $>4$), and withstand robustness checks standard to high-resolution spectroscopy literature. As an additional check, as well as to explore the efficacy of airmass- and PCA-based detrending methods on different datasets, we apply both detrending methods to the 2.3$\mu$m and 3.2$\mu$m observations of HD 189733 b. Both methods require optimizing several hyperparameters. A detection which is robust to both methods is unlikely to arise from overfitting.

\begin{figure*}
	\includegraphics[width=0.50\columnwidth,trim={9cm 0 3.5 0},clip]{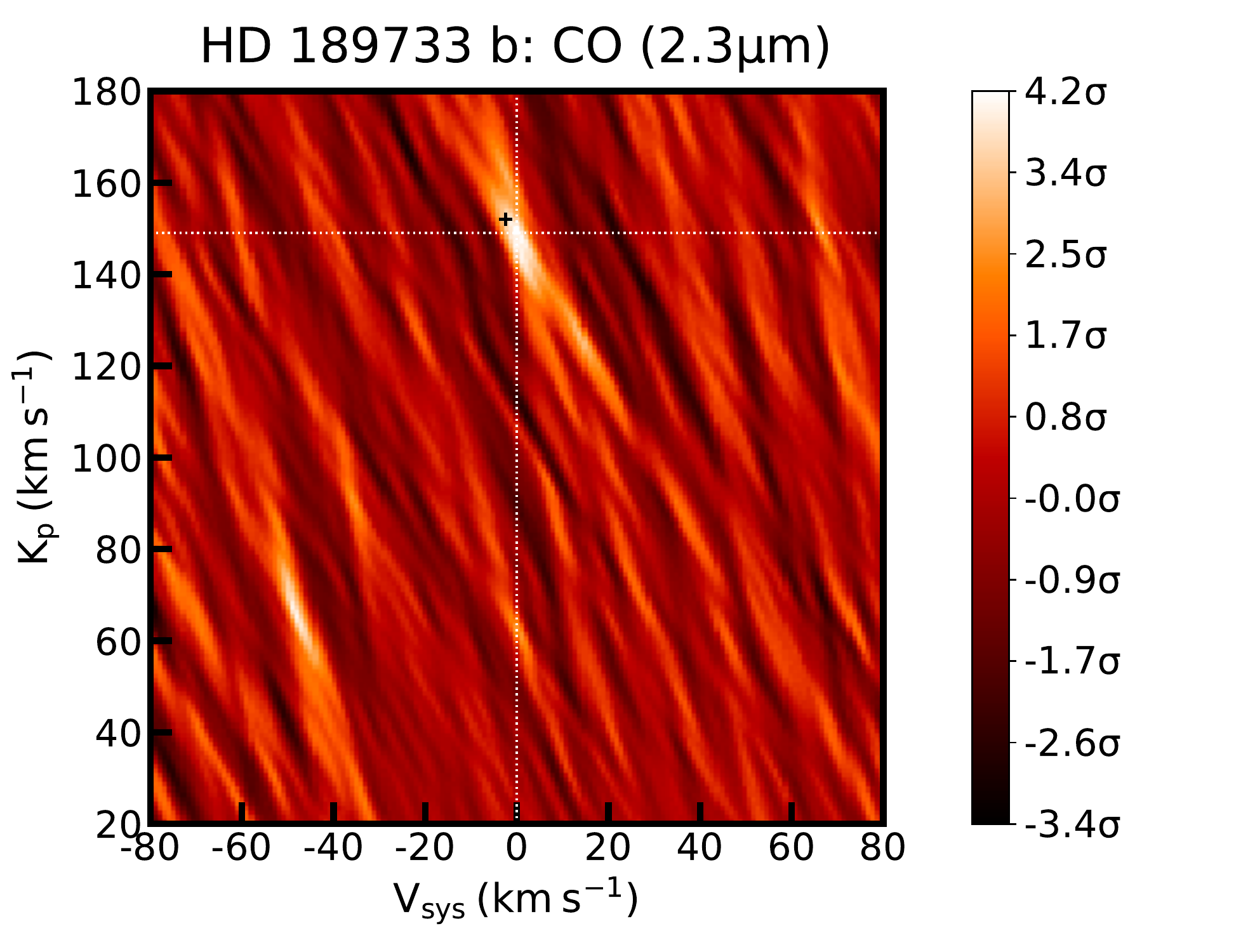}
	\includegraphics[width=0.50\columnwidth,trim={9cm 0 3.5 0},clip]{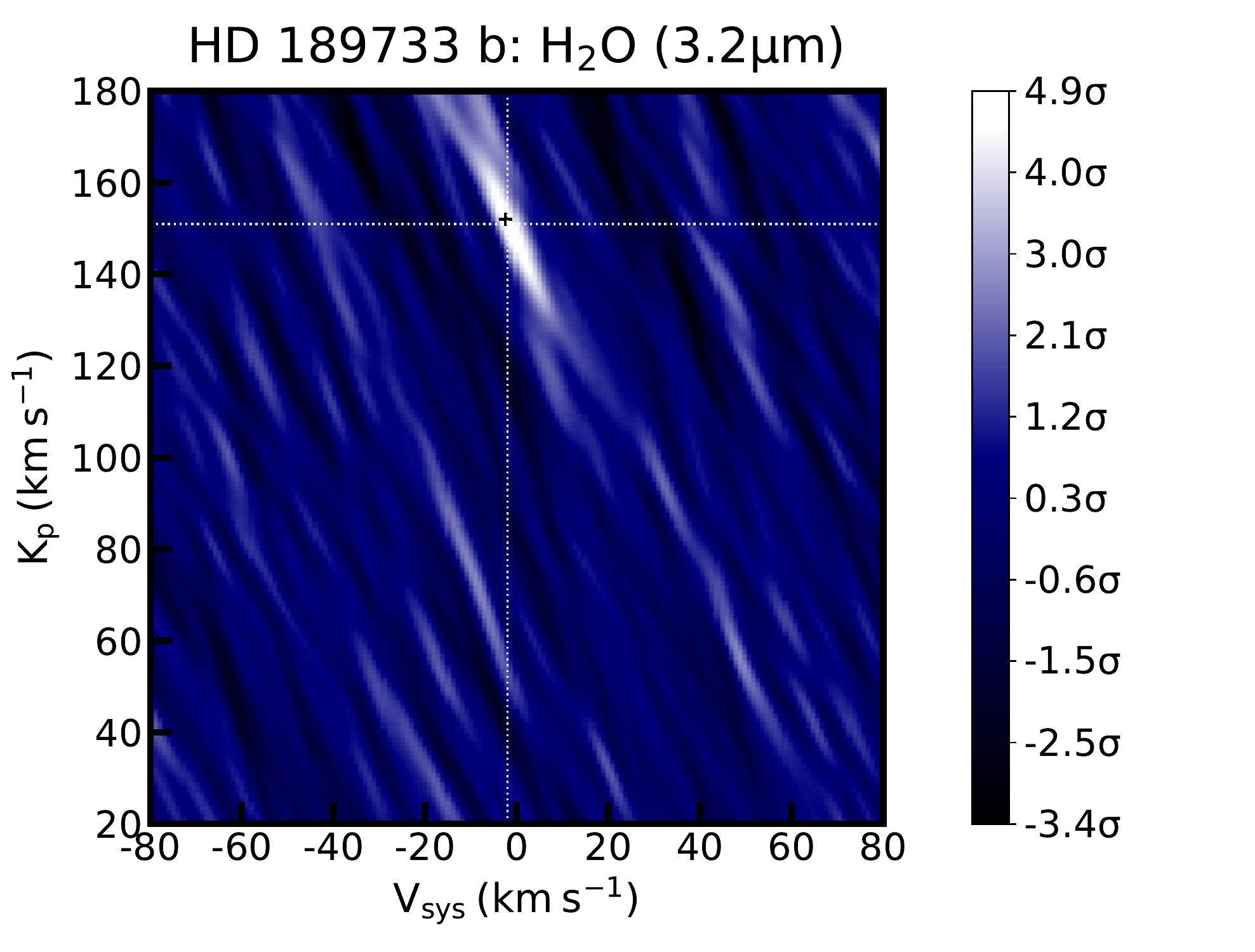}
	\includegraphics[width=0.50\columnwidth,trim={9cm 0 3.5 0},clip]{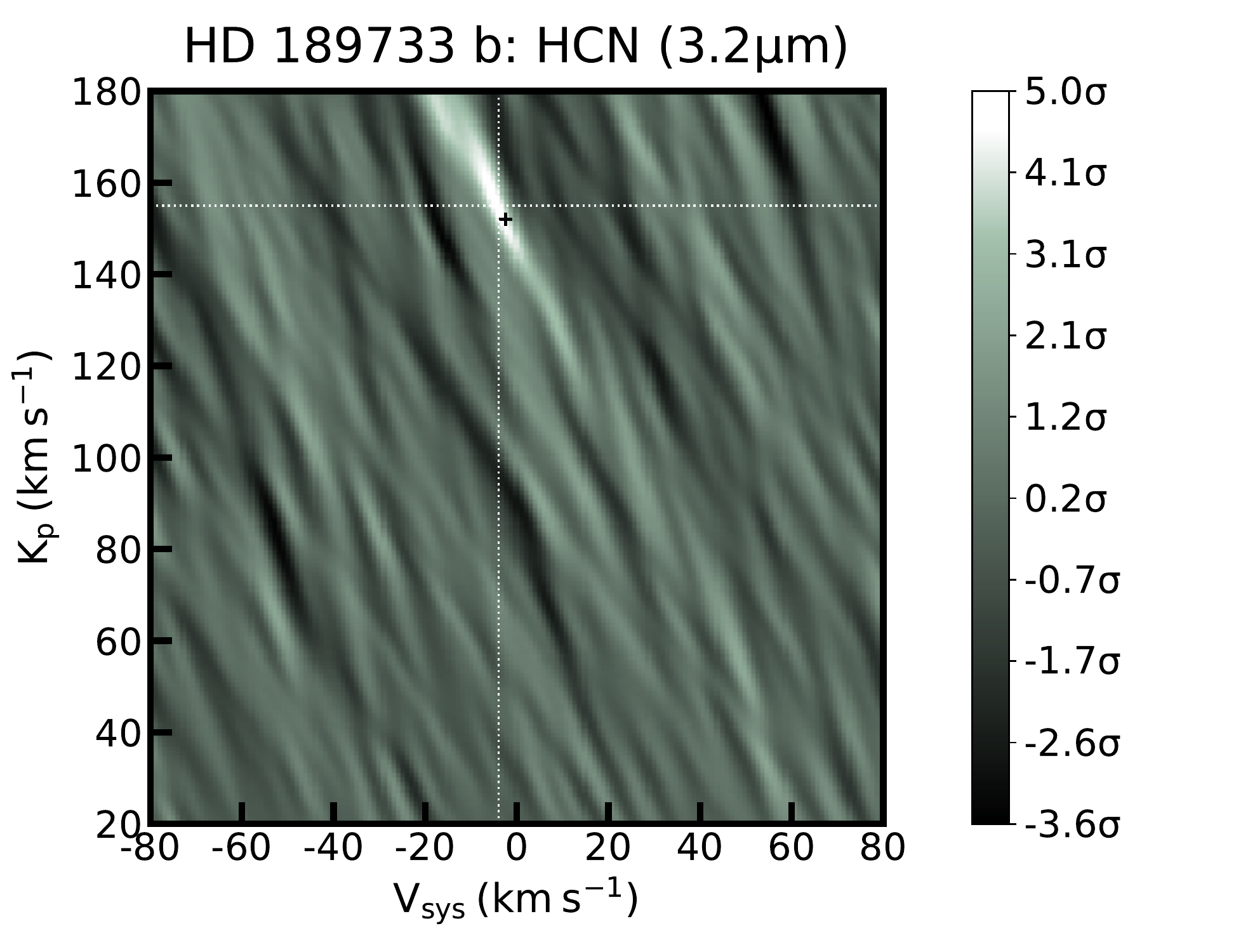}
    \includegraphics[width=0.50\columnwidth,trim={9cm 0 3.5 0},clip]{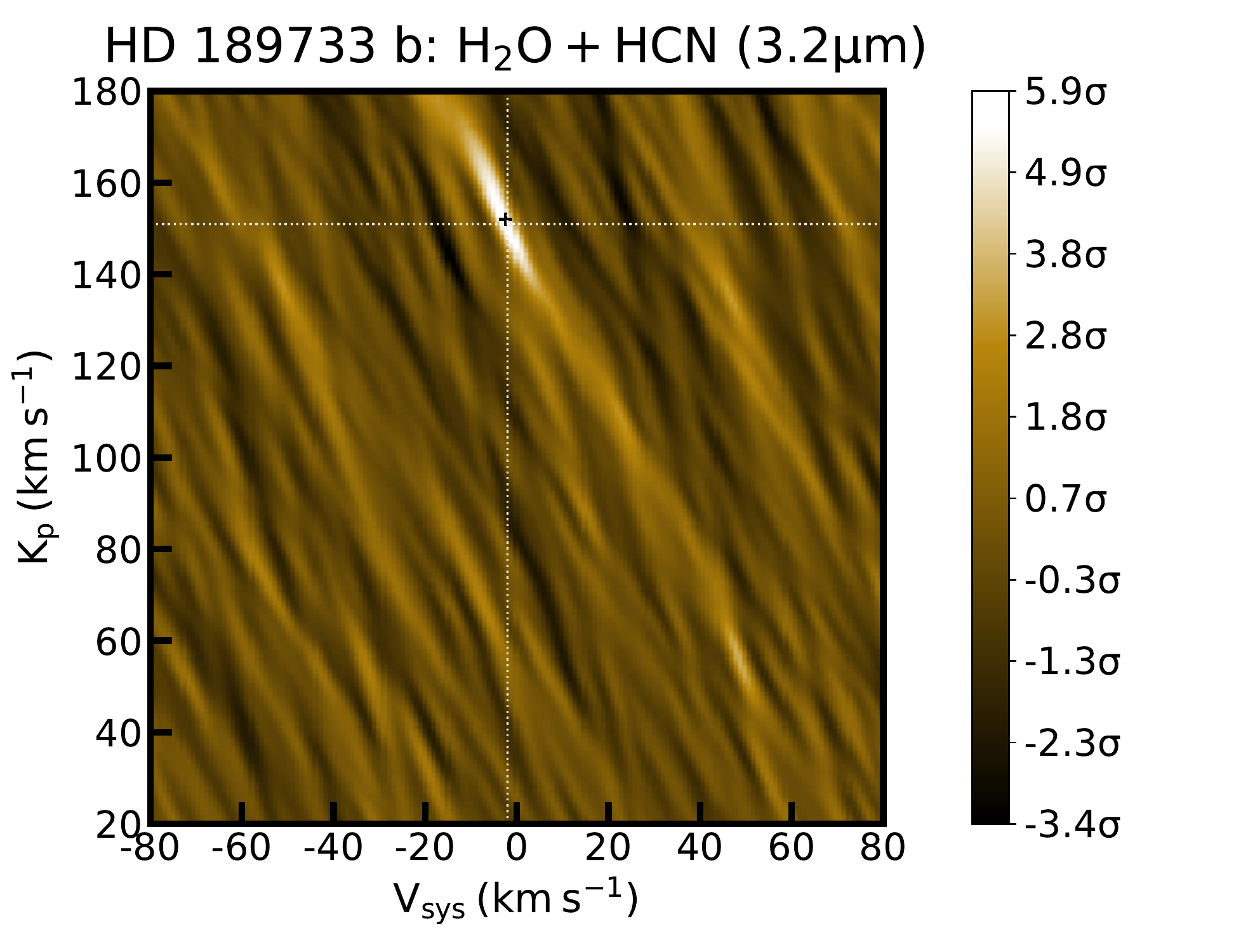}
    \\
	\includegraphics[width=0.50\columnwidth,trim={9cm 0 3.5 0},clip]{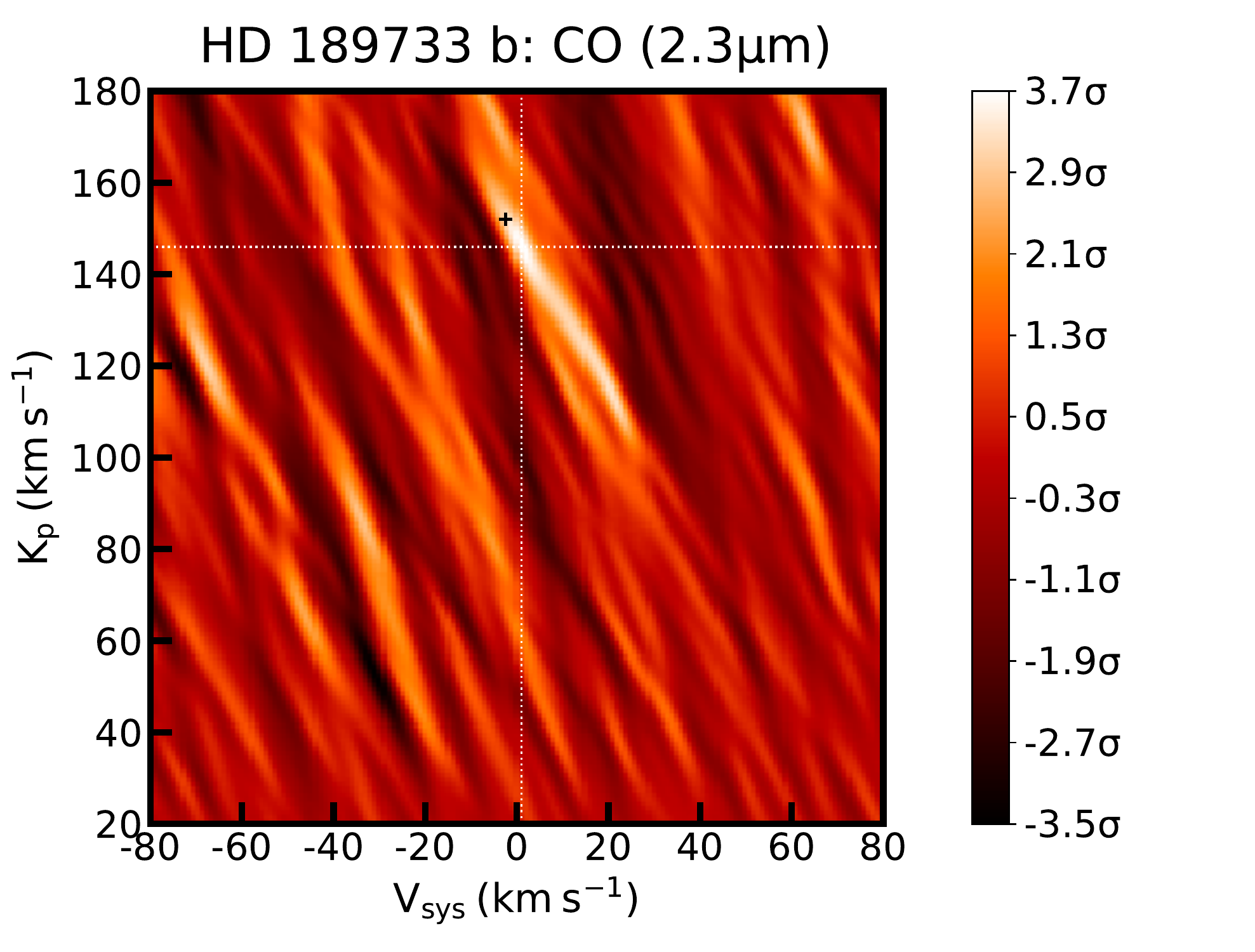}
	\includegraphics[width=0.50\columnwidth,trim={9cm 0 3.5 0},clip]{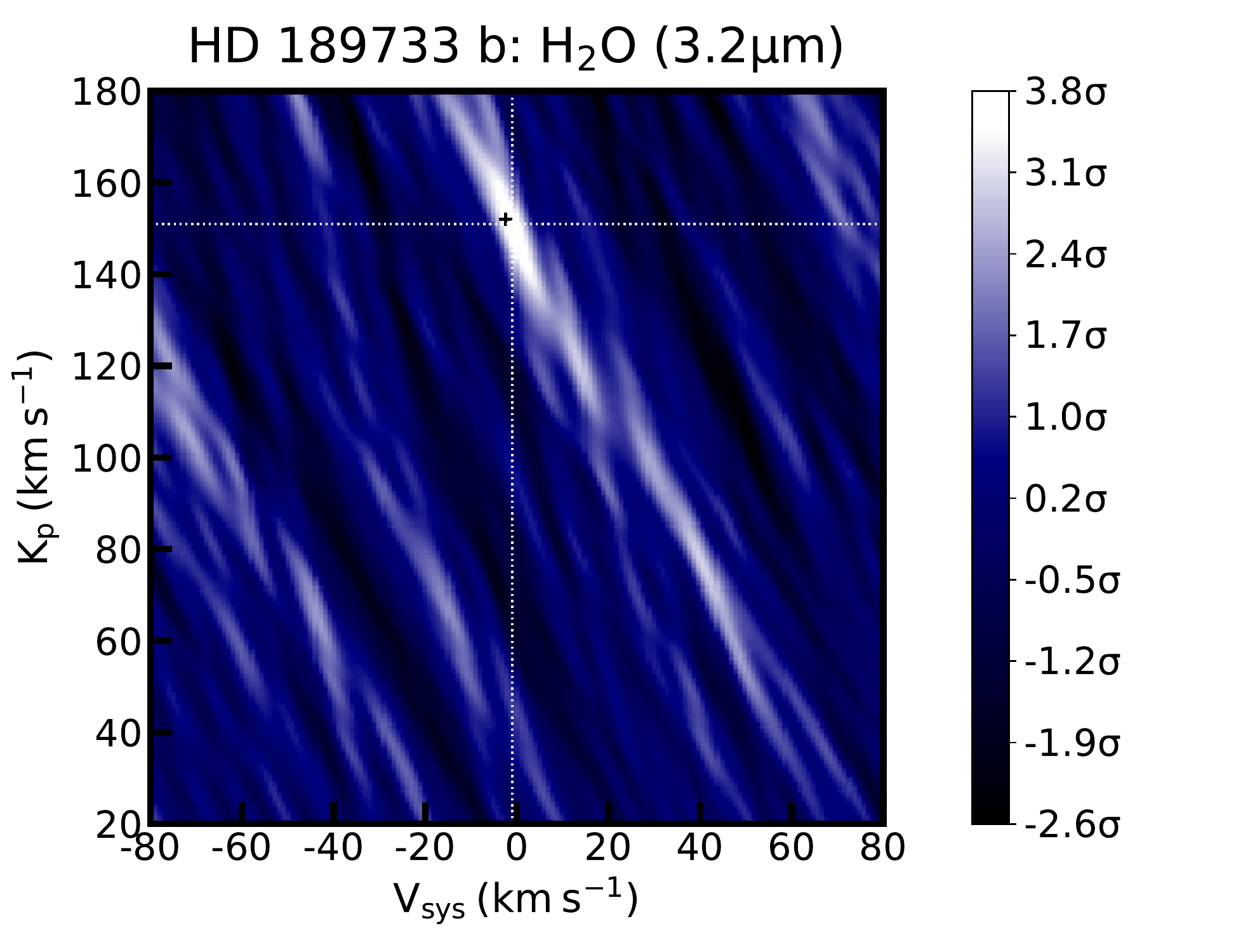}
	\includegraphics[width=0.50\columnwidth,trim={9cm 0 3.5 0},clip]{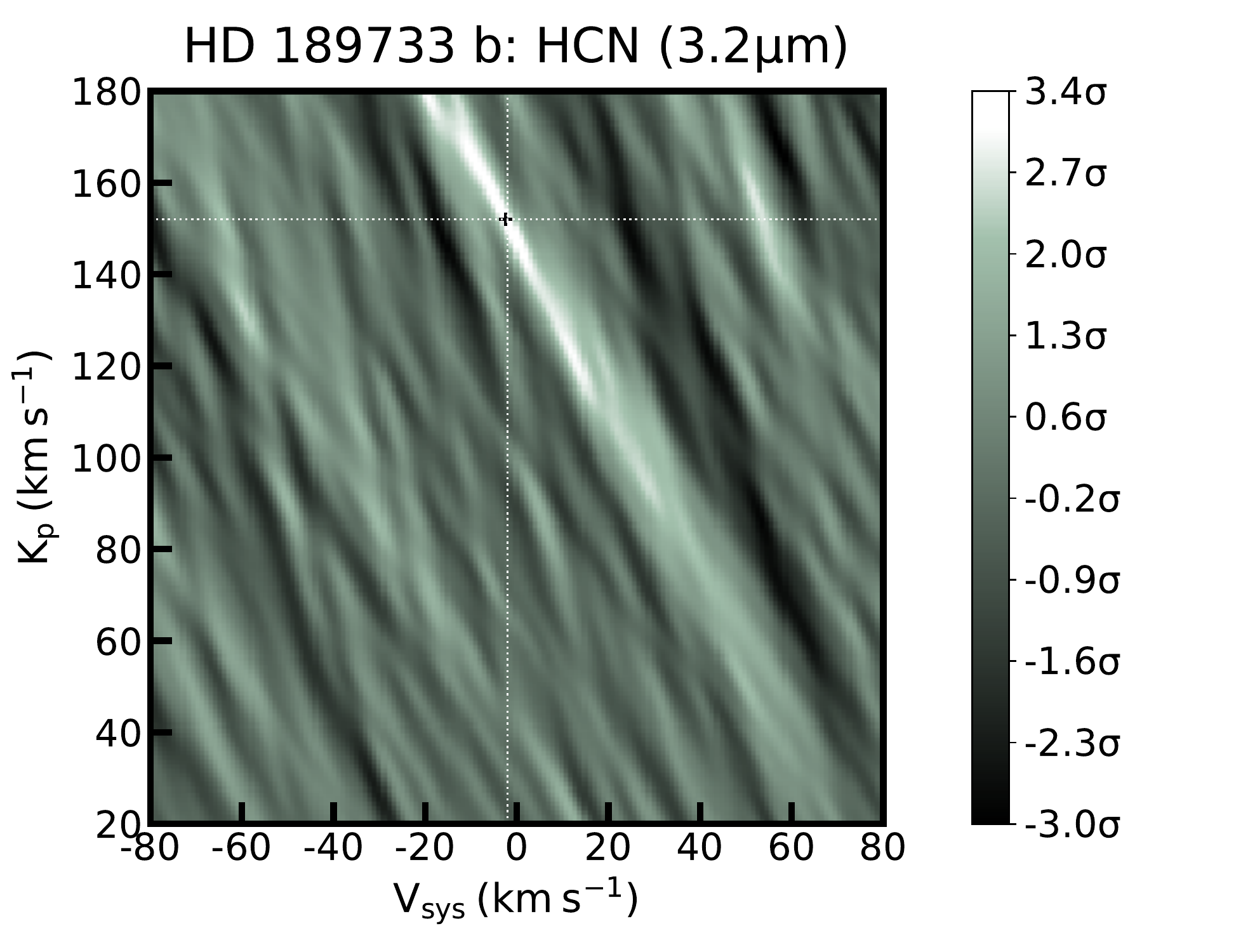}
    \includegraphics[width=0.50\columnwidth,trim={9cm 0 3.5 0},clip]{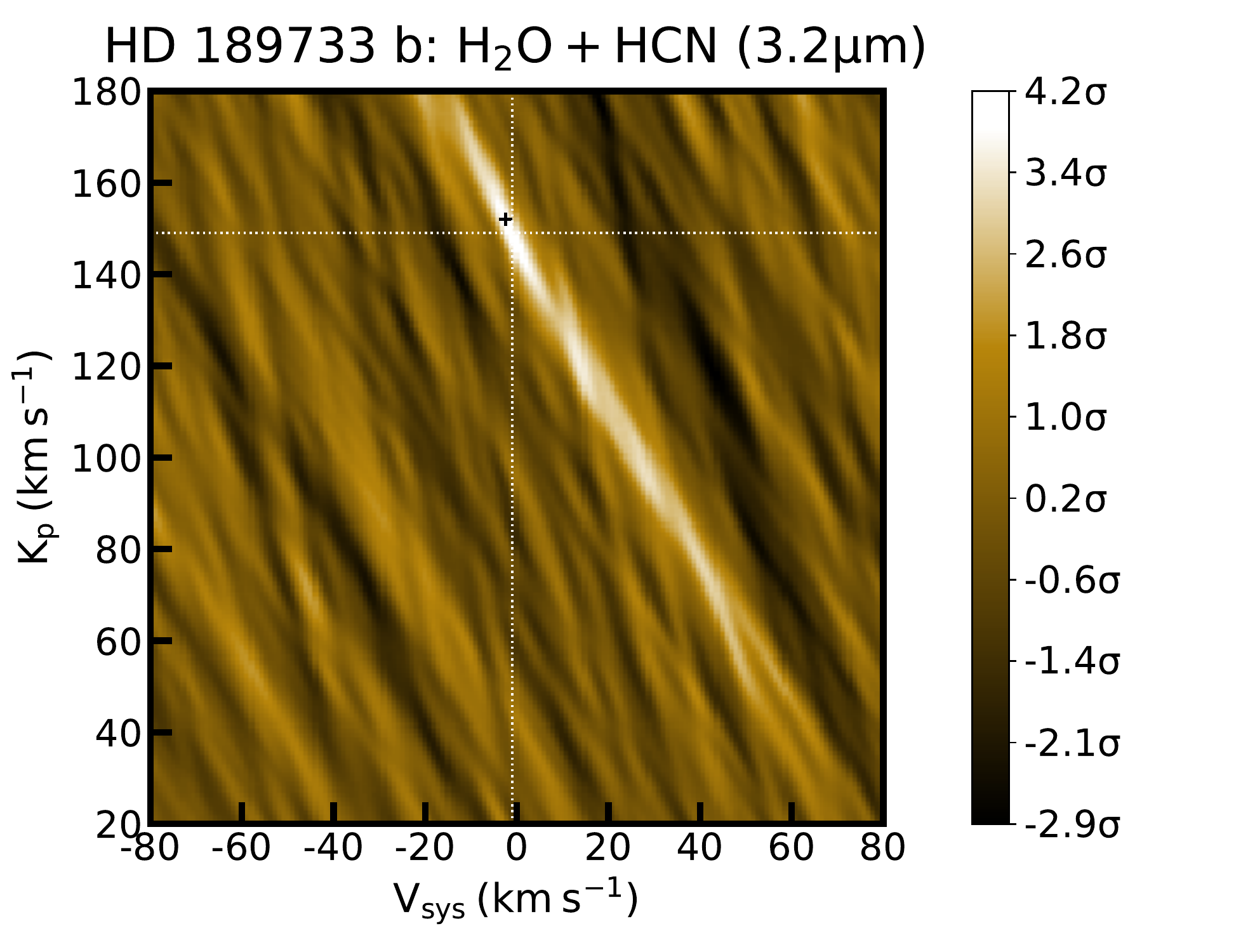}    
    \caption{
    Molecular detection significances for HD 189733 b using the two detrending methods. From left to right: detection significances of CO in the 2.3$\mu$m band, and of H$_2$O, HCN, and combined H$_2$O + HCN in the 3.2$\mu$m band. The black plus indicates the expected planetary $K_p$ and $V_{sys}$. The top row of detections are made using the SYSREM detrending method. The bottom row of detections use airmass fitting.}
    \label{fig:hd_comp}
\end{figure*}

Previous high-resolution studies adopt several variants of airmass-based detrending. Indeed, detrending must be performed carefully in order to remove nuisance signals, but not significantly degrade the weak planet signal. The optimal methodology depends on several factors, such as the strength of the planet signal compared to the noise level, the amount of data, and the presence of telluric and systematic effects on individual detectors. Our confirmation of CO in Tau Boo b in Section~\ref{sec3} validates the airmass-based detrending procedure outlined therein. We apply the procedure to 2.3$\mu$m observations of HD 189733 b spanning a single night. We determine the appropriate amount of masking by injecting a CO planetary model, and maximizing the significance of its recovery over $p_v$ and $p_m$. Cross-correlation with a CO template yields evidence for absorption at a significance of 3.2$\sigma$. The detection peak is at a location consistent with the expected planet $K_p$ and $V_{sys}$ \citep{triaud2009,bouchy2005}, within uncertainty. However, spurious positive and negative contours obtain comparable amplitudes ($\sim 3 \sigma$) as the peak significance; thus we cannot conclusively classify the peak as a detection. 

In their analysis of HD 179949 b observations, \citet{brogi2014} do not apply a high-pass filter, and renormalize each column of the data by its variance prior to cross-correlation. Using this alternative approach on the 2.3$\mu$m observations of HD 189733 b yields an improved detection significance of $3.4\sigma$. 
\citet{brogi2016} remove a quadratic fit with airmass, use a high pass filter, and renormalize each column of the data by its variance. They do not sample residuals for higher-order detrending. This approach yields a detection significance of $3.3\sigma$  for CO in HD 189733 b. The noise patterns in the resulting CCF and velocity-space detection significances are similar for each method, suggesting they all accomplish the same task of detrending resuiduals and isolating the planet signal. Their level of success depends on fine-tuning for the dataset at hand. We obtain the highest significance CO detection (3.7$\sigma$) by performing a linear fit with airmass, sampling 9 columns for higher-order detrending, applying a high-pass filter, and normalizing each column by its standard deviation. We conduct no further optimizations. In order to avoid degrading the planet signal, some previous studies mask columns that are expected to contain strong absorption features from parts of the detrending process \citep{snellen2010,schwarz2015}. It is also common to either exclude detectors (e.g. because of odd-even interference) \citep{brogi2012,birkby2013,birkby2017,brogi2016,brogi2014} or weight contributions from individual detector CCFs \citep{brogi2017}. In this study, however, we focus strictly on methods which use all available data, and treat the data equally with minimal prior knowledge of the planetary system.

Figure \ref{fig:hd_comp} shows summed-CCF detection significance plots from: cross-correlation with H$_2$O, HCN, and combined H$_2$O $+$ HCN templates for the 3.2$\mu$m dataset, and cross-correlation with a CO template for the 2.3$\mu$m dataset. We use both SYSREM detrending outlined in the previous section, and the airmass detrending method that yields the highest CO detection significance described above. The results from the two methods are shown in the top and bottom rows of Figure \ref{fig:hd_comp} respectively. For both wavelength regimes and detrending methods, we identify the optimal level of masking via model injection and recovery. In all cases, we obtain peak detection significances at $K_p$ and $V_{sys}$ consistent with the known values. In both wavelength regimes, SYSREM consistently outperforms airmass-based detrending, with more than $+1\sigma$ higher detection significances. This performance is most likely related to the severe telluric contamination. Since SYSREM is not restricted to specific columns of the data, it can more easily model systematics, environmental variations and telluric variation. Indeed, \citet{dekok2013} and \citet{birkby2017} mention several possible sources of trends, including airmass, Adaptive Optics Strehl Ratio, and temperature. Airmass detrending performs slightly better in the 2.3$\mu$m since it is less contaminated by tellurics

\subsection{Sensitivity to SYSREM Parameters}
As outlined above, detrending the data using SYSREM requires several hyper-parameters for each detector, namely the number of iterations as well as flux and variance cut offs for telluric masking. We explore the sensitivity of the analysis to these hyperparameters below.

The optimal number of SYSREM iterations, N, is set for each detector individually by recovery of an injected model, following similar studies that use SYSREM \citep{birkby2013,birkby2017,nugroho2017}. This is done since too few SYSREM iterations would mean significant telluric contamination remains whilst too many iterations would degrade the planetary signal. While the component removed in the first iteration resembles airmass, components removed by further iterations can pertain to subtle, higher-order trends which are less understood. Some examples of possible trends are shown in \citet{dekok2013} and \citet{birkby2017}. We show the recovered strength of the injected model as a function of N for all 4 detectors in Figure \ref{fig:sysdet}.

\begin{figure}
\centering
\includegraphics[width=\linewidth,trim={1cm 0 2cm 0},clip]{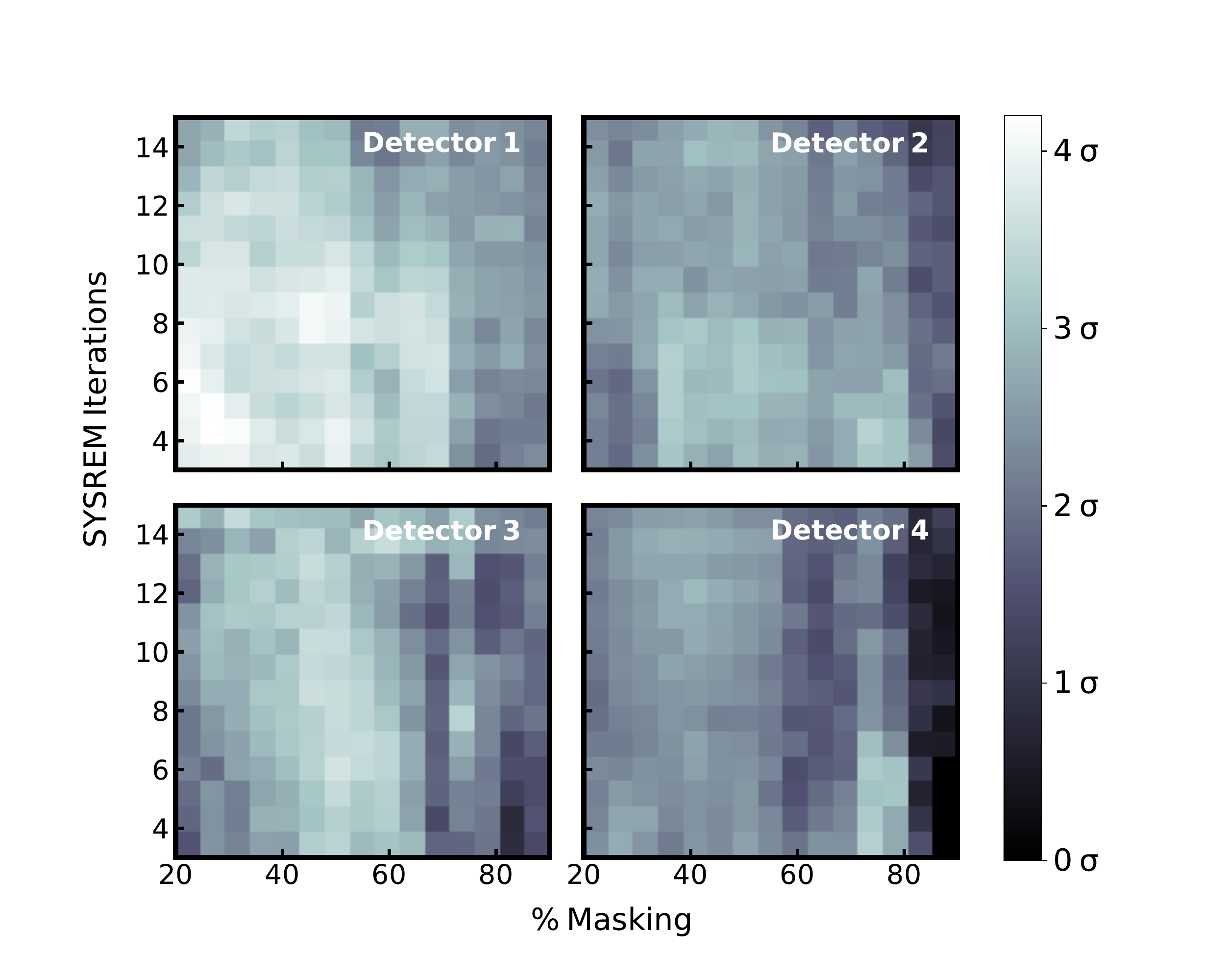}
\caption{Recovered detection SNR for each detector as a function of applied SYSREM iterations and the percentage of masking. For ease of plotting the percentage of masking is the sum of the percentages of highest variance and lowest mean columns used and arbitrarily divided between the two criteria.  The recovery is based on injection of an HCN model at $1\times$ nominal strength. The dataset of HD 189733 b in the 3.2$\mu$m band is used here for illustration.
}
\label{fig:sysopt}
\end{figure}

Additionally, the optimal number of SYSREM iterations has a strong dependence on the masking applied to telluric regions. We find the recovered signal varies with the amount masking applied to telluric regions in each detector. The masking is set through considering the lowest mean columns and highest variance columns. A percentage of each are masked with the percentage set either manually or determined by maximizing the recovery of an injected model. The latter is motivated by the interdependence between the number of SYSREM iterations and the masking applied. Both this interdependence and fundamental dependence of the results on masking are shown in Figure~\ref{fig:sysopt} which shows the recovery strength of an injected signal for different masking percentages and SYSREM iterations. The interdependence of masking and SYSREM is expected given that the principal components of the data array will change significantly when more/less columns with telluric contamination are masked. Using a quantitative masking criteria is very powerful given its reproducibility and the ability to calibrate the hyperparameters through the recovery of an injected model. Additionally the quantitative criteria allows us to show the strong dependence of search results on masking and as such highlight it as a key step.

\subsection{Spurious Detections in Velocity-Space}
\begin{figure*}
\centering
\includegraphics[width=0.8\linewidth,trim={1cm 1.6cm 2cm 1cm},clip]{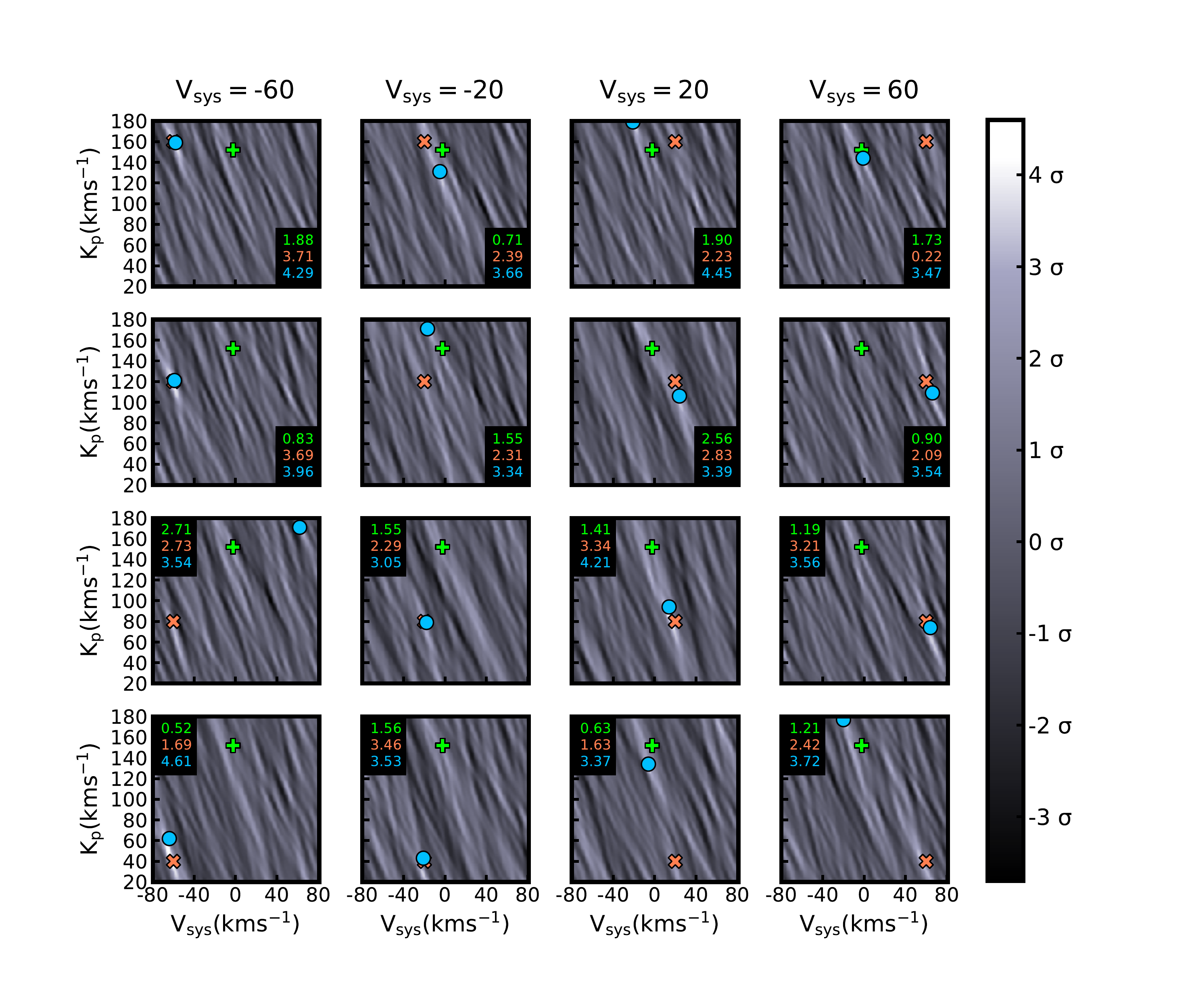}
\caption{Detection significances with detrending using SYSREM and optimized to different $K_p$ and $V_{sys}$. Cross-correlation is performed with the HCN spectral template. In each panel, SYSREM is optimized to a specific $K_p$ and $V_{sys}$ as denoted by the location of the coral X, which set the detrending parameters for that panel. The planet location is shown as the green + and the location of the peak significance in each panel is marked by a blue circle. The legend in each panel shows the values corresponding to each marker, by color.
}
\label{fig:inj_grid_plot_sysrem}
\end{figure*}

\begin{figure*}
\centering
\includegraphics[width=0.8\linewidth,trim={1cm 1.6cm 2cm 1cm},clip]{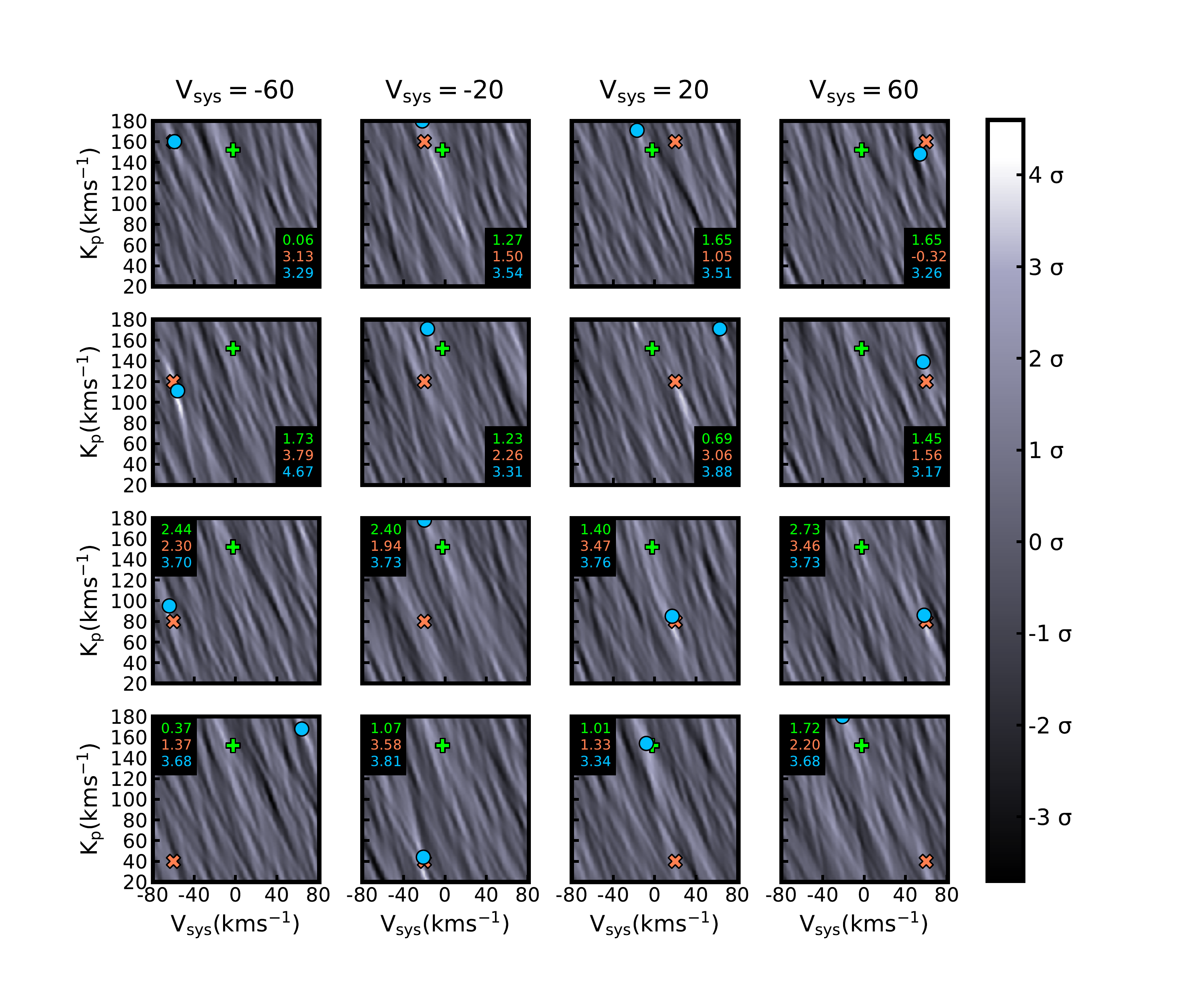}
\caption{Detection significances with detrending using airmass fitting and optimized to different $K_p$ and $V_{sys}$. Cross-correlation is performed with the HCN spectral template. In each panel, airmass fitting is optimized to a specific $K_p$ and $V_{sys}$ as denoted by the location of the coral X, which set the detrending parameters for that panel. The planet location is shown as the green + and the location of the peak significance in each panel is marked by a blue circle. The legend in each panel shows the values corresponding to each marker, by color.
}
\label{fig:inj_grid_plot_airmass}
\end{figure*}

In optimizing the detrending parameters, it is important to assess the possibility of false-positives. In determining the number of SYSREM iterations through optimizing the recovery of an injected model we must choose a $K_p$ and $V_{sys}$ for the injected model. Following methods common to the literature \citep{birkby2013,birkby2017,nugroho2017} we inject the model at the known planetary $K_p$ and $V_{sys}$ to maximize the chance of finding a planetary signal. Here we optimize the recovery of models injected at other $K_p$ and $V_{sys}$ locations. We then use the hyperparameters determined for various $K_p$ and $V_{sys}$ locations on the real data without injection to check whether we are able to get significant peaks at $K_p$ and $V_{sys}$ locations where there should be none. We investigate this dependency to check for potential false positives and the results are shown in Figure~\ref{fig:inj_grid_plot_sysrem}. We repeat this test using airmass-based detrending as shown in Figure~\ref{fig:inj_grid_plot_airmass}.

In about $30\%$ of cases, optimizing SYSREM to another location yields a false positive detection; these are cases in which we find $3-4.5\sigma$ peak significance, and the peak is at a location consistent within a few km s$^{-1}$ of the optimization location. Such signals could potentially result from an over fitting with the number of hyperparameters required in the detrending process with SYSREM and masking for each detector. Optimizing airmass to other locations yields the same fraction of false-positives. When optimizing and detrending with SYSREM, noisy peaks reach as high as $4.6\sigma$, and are typically around $4.0\sigma$. When detrending with airmass, noisy peaks reach as high as $4.7\sigma$, and are typically around $3.5\sigma$. In most cases, the peak contour in the plots does not correspond with the injection location. However, there are several cases in which the false positive peak matches the optimization location. These cases are the same when detrending with SYSREM and with airmass, suggesting that masking is the dominant parameter causing velocity-space dependence. While these findings are alarming at first glance, we note that previous works \citep{brogi2013,birkby2013,birkby2017,dekok2013} often obtain anti-correlations in velocity space at $\lesssim -4\sigma$. Since the detection significances are derived from the sum of Gaussian CCFs, one can expect positive spurious correlations of the same magnitude, especially if optimized to spurious local maxima.

Regardless of location, we maintain a consistently positive (approximately $1-3\sigma$) local maximum at the planetary location. Additionally, the detection peaks for CO, H$_2$O and HCN presented earlier directly correspond to the planetary $K_p$ and $V_{sys}$. The H$_2$O and HCN detections made with SYSREM are also of slightly higher significance than the spurious peaks resulting from this test. Together, these findings boost confidence in the robustness of our SYSREM-based molecular detections. False-positives from airmass detrending can surpass the detection significances of our molecular detections. This is probably because airmass detrending is not as well-suited as SYSREM for detrending the 3.2$\mu$m dataset \citep{birkby2013,birkby2017,birkby2018}.

\begin{figure*}
\centering
\includegraphics[width=0.95\linewidth,trim={3.5cm 1.6cm 3.0cm 0.4cm},clip]{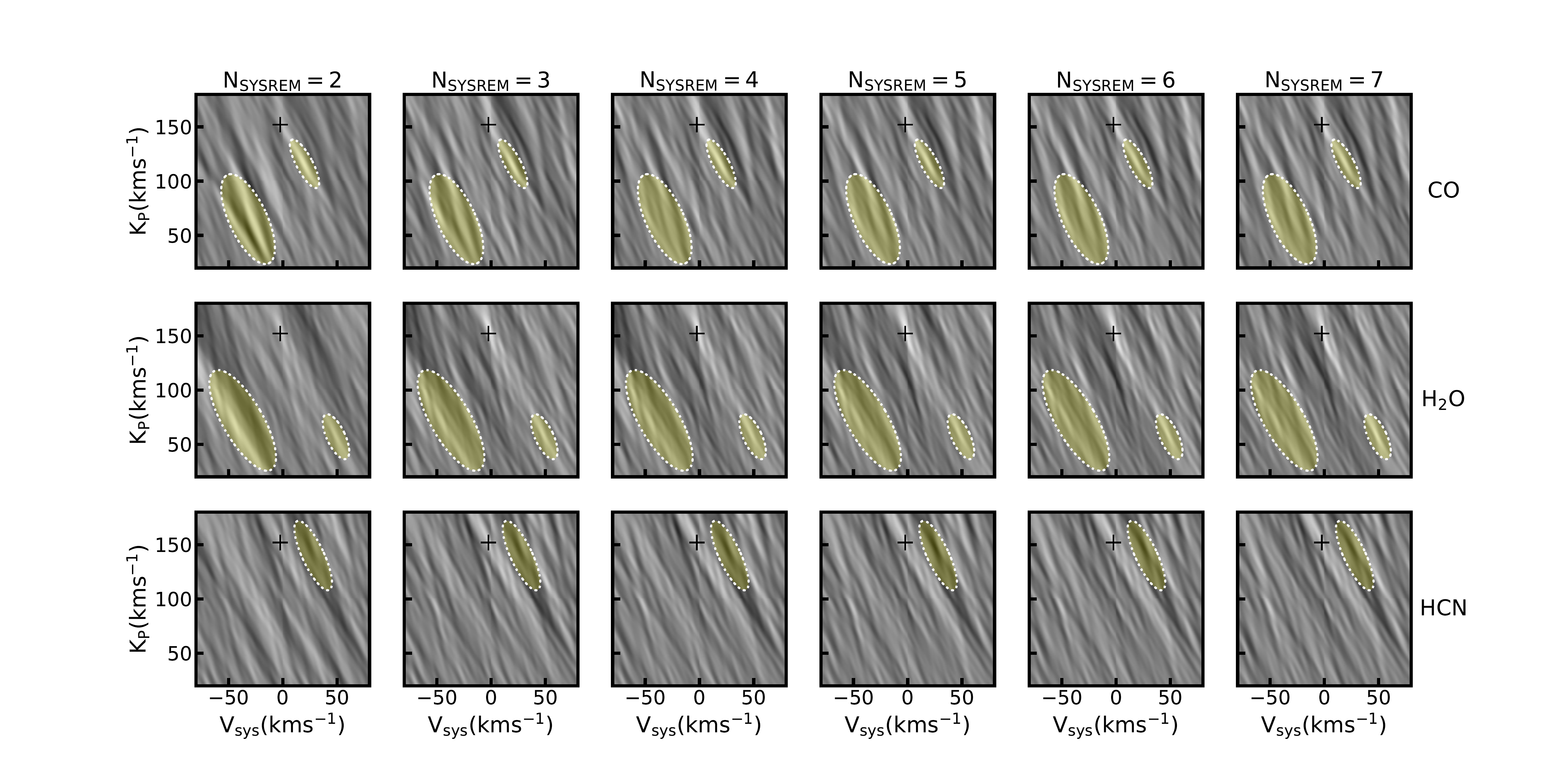}
\caption{The propagation of residuals with increasing SYSREM iterations, for each molecular template. Detection significances are obtained using $N_{SYSREM}$ iterations for all four detectors. The black plus marks the expected location of the planetary signal. Shaded regions mark the strongest residuals in the early detrending stages and their evolution with increasing SYSREM iterations is shows in different columns for each molecule. For example, a strong residual is seen initially in the bottom left part of the CO panel, as shown in the larger ellipse, but is removed with subsequent detrending iterations. This may be similar to the stellar residual predicted by \citet{dekok2013} On the other hand, other residuals can persist the SYSREM iterations, as shown in the smaller ellipses, but are generally weaker than the planetary signal. These latter residuals may be responsible for false positives in some conditions.
}
\label{fig:sys_res}
\end{figure*}

These findings highlight the importance of making iteration and masking parameters explicit and quantifiable so that such checks can be performed. When detrending is optimized to other locations than the planet, noisy peaks can reach high SNR levels ($\gtrsim 4\sigma$). This fact suggests that, when optimizing to the planetary location, the significance of the molecular detection might be overestimated. The origin of spurious peaks is difficult to determine, but potential sources include residuals from stellar spectral features and telluric absorption not removed by detrending. For example, previous studies indicate that CO stellar spectral features can lead to strong cross-correlation residuals \citep{brogi2013,dekok2013}, albeit at different $K_p$ and $V_{sys}$ than the planetary signal. We perform a test in which we fix the number of SYSREM iterations applied to each detector, and visualize the persistence or removal of strong residuals with additional iterations (Figure~\ref{fig:sys_res}). 

We identify a strong residual from cross-correlation with CO at $K_p \sim 80$ km s$^{-1}$ and $V_{sys} \sim -40$ km s$^{-1}$, consistent with predictions in \citet{dekok2013}. Successive SYSREM iterations quickly remove it, suggesting stellar contamination is unlikely to be responsible for persistent, spurious signals. Other, smaller residuals from cross-correlation with CO, H$_2$O and HCN can occasionally persist. In the HCN case, the spurious peaks span a range of $K_p$ and $V_{sys}$ that may not necessarily correspond to physically plausible velocities. Such false positives are most likely due to overfitting (detrending and masking optimization) to noise in the CCF, which is set by the data quality/quantity and cross-correlation template.

\subsection{Metrics for Detection Significance}

\begin{figure}
\centering
\includegraphics[width=0.8\linewidth,trim={0 4cm 1.5cm 5.2cm},clip]{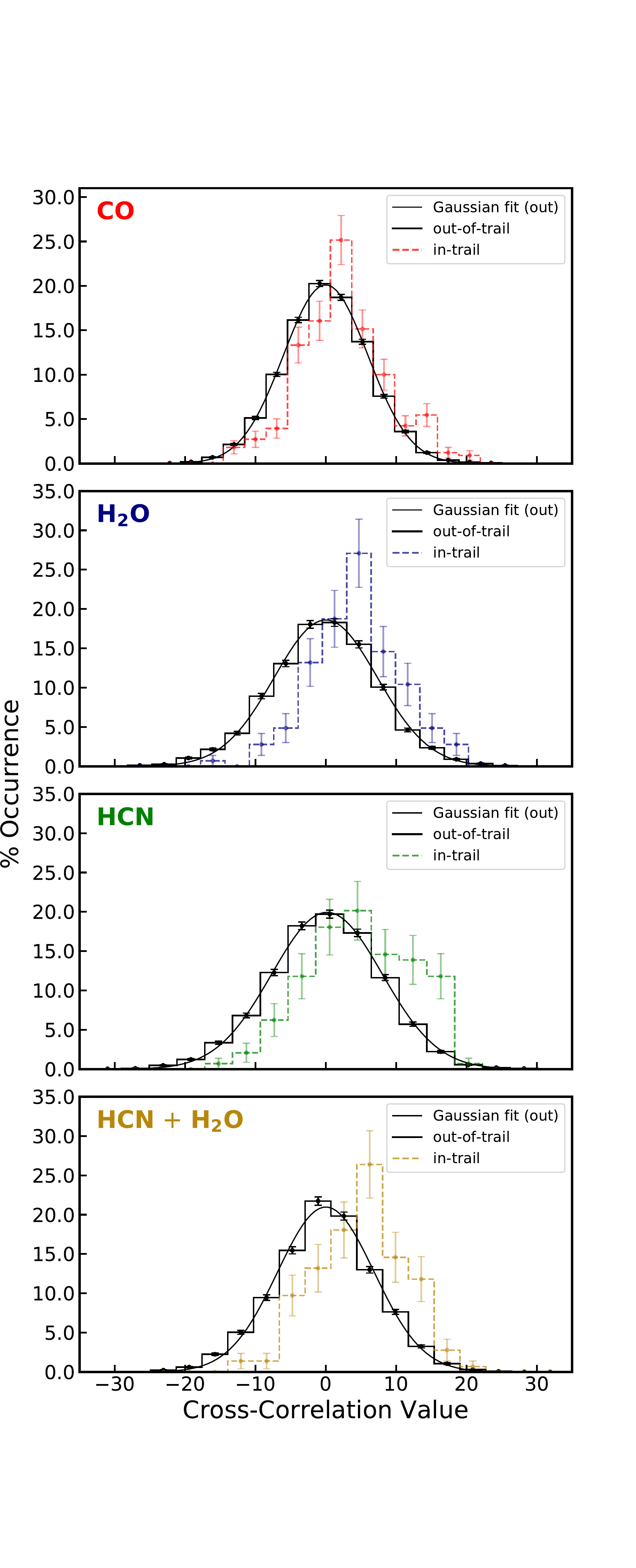}
\caption{Distributions of `in-trail' and `out-of-trail' Cross-Correlation Function values. These are shown for correlation with CO, H$_2$O, HCN and H$_2$O + HCN model templates. The `out-of-trail' best-fit Gaussian is plotted in each case. Error bars are the square root of the bin population size. The `in-trail' distributions are consistently shifted toward higher values, and the `out-of-trail' distributions are highly Gaussian. For each template, Welch $T$-tests reject consistency between the two distributions at a $>6\sigma$.
}
\label{fig:dist_plot}
\end{figure}

The two different metrics typically used to quantify the detection significance are an SNR metric \citep{brogi2012,brogi2013,brogi2014,brogi2018} and a Welch $T$-test \citep{birkby2017,nugroho2017,brogi2018}. We denote in-trail pixels as those which lie in a 3-pixel wide window centered on the dark trail in Figure~\ref{fig:ccf_plot}, while out of trail pixels are the remaining ones. In the Welch T-test, the distributions of in trail and out-of-trail pixels are compared to determine whether they are drawn from different Gaussian distributions. The SNR metric however is simply the array of summed CCFs normalised by their standard deviation. Typically we find the Welch $T$-test returns a higher detection significance with our results from SYSREM yielding $>6\sigma$ for each molecular template. The in-trail and out-of-trail distributions are shown in Figure~\ref{fig:dist_plot} with all plots suggesting the distributions of in-trail pixels have greater means than those of the highly Gaussian out-of-trail distributions. These higher significances suggest the Welch T-test could be more vulnerable than the SNR metric to potentially overestimating the confidences of detections.

We investigate sensitivity of the Welch T-test in two ways. First, we sum a varying number of pixels (1, 3, 5, and 7) centered on the planetary signal. Second, we repeat the analysis using different velocity space resolutions (0.5, 1.0, and 1.5 km s$^{-1}$). The detection significances increase dramatically with higher velocity space resolution and more sampling of in-trail pixels. In this way, we artificially increase the sample sizes of both the in-trail and out-of trail distributions, and thus overestimate the confidence predicted by the Welch T-test. An ideal metric would be robust to oversampling pixels in the two distributions. We note that the SNR metric does depend on the width of the in-trail. Sampling too many pixels draws values from the noise distribution and smears out the signal. Sampling fewer pixels can help ensure they are all drawn from the in-trail, but with the risk of losing part of the signal. We find 3 pixels strikes a balance that maximizes the significance of the planetary signal.
We also investigate the SNR metric as the standard deviation of the $K_p$--$V_{sys}$ plot shows some non random variation with $K_{p}$. We investigate this variation and find that beyond $K_p\approx100$km s$^{-1}$ the standard deviation is approximately constant whilst for smaller $K_{p}$ the standard deviation is smaller. As such using the standard deviation of the entire $K_p$ space overestimates the detection significance for higher $K_p$ hence we recalculate significances by calculating the standard deviation for $K_{p}>100$km s$^{-1}$. This results in a small decrease of up to $\sim0.5 \sigma$ in the quoted detection significances. Previous studies take various approaches with some more recent works using the standard deviation for each $K_p$ and others using the standard deviation from the whole $K_p$ space \citep{brogi2012,brogi2014,brogi2016, dekok2013, birkby2013}.

Given the Gaussianity of the CCFs, the negative contours from anti-correlation up to the $4\sigma$ level imply the presence of false positive correlations at the same level. This is another potential flag for the overestimation of detection significances, such anti-correlation signals are commonly seen in the literature and may imply that the noise distribution is not well characterized by the metric used. However, strong anti-correlations could also potentially be attributed to the intrinsic autocorrelation properties of a species.

High-resolution spectroscopy may require a different metric to determine confidences in molecular detections. The ideal significance metric is one which captures how unlikely that significance is to arise by chance. While SNR and Welch T-test are statistically reasonable metrics, they might not reflect the true confidence in a molecular detection. We see that high-significance peaks based on these metrics are not uncommon when one optimizes analysis methods to a desired location. 

\section{Summary and Discussion}\label{sec5}

We present a systematic analysis of detrending methods in high-resolution spectroscopy, and their effect on chemical detections in exoplanetary atmospheres. In particular, we investigate detrending with two methods: the PCA-based algorithm SYSREM; and airmass fitting with subsequent sampling of higher-order trends. Both methods entail fine-tuning hyperparameters in order to obtain high-significance molecular detections. We explore the dependence of detection significances on these hyperparameters, using the hot Jupiter HD 189733 b as a case study for this purpose. We confirm previous detections of CO \citep{dekok2013} and H$_2$O \citep{birkby2013} in the planet using varied techniques, and report a new possible detection of HCN.

We perform several tests to address the robustness of these detections. Additionally, we present a detailed analysis of the sensitivity of the detection significances to the hyperparameters of detrending. We find strong dependence on the number of SYSREM iterations and the amount of masking of tellurics. We also find sensitivity to masking and residual sampling in airmass-based detrending methods. By optimizing the detrending procedure to different $K_p$ and $V_{sys}$, one can obtain high SNR spurious peaks, which raises concerns over false-positives. However, the detection of species using both detrending procedures can help ensure robustness. Additionally, maintaining a relatively high SNR at the planetary orbital parameters during optimization at other locations can also provide evidence against false positives.

Many previous high-resolution studies use SYSREM or airmass-based detrending \citep{brogi2012,birkby2013}. The analysis typically involves some form of masking bad columns of the data. Additionally, airmass-detrending methods typically require sampling additional columns from the data to perform higher-order detrending. Between various studies, assumptions tend to be made on a case-by-case basis depending on the dataset, seldom quantified in a reproducible manner. We have shown, however, that varying the amount of masking and the sampling of additional columns can lead to strong spurious signals at levels of $\gtrsim 4\sigma$. In the case of SYSREM, optimization to the planetary $K_p$ and $V_{sys}$ is typically performed to determine the number of iterations. Masking is also performed; although the criteria for classifying bad columns remains arbitrary. We have shown that optimization has strong dependence on the masking stage. Additionally, we have investigated optimization to non-planetary $K_p$ and $V_{sys}$, and obtain spurious signals at levels of $\gtrsim 4\sigma$.

Our analysis demonstrates the need for consistent and robust detrending methods, in addition to quantifiable and reproducible constraints on hyperparameters such as masking, sampling and iterations. One must take caution so as to not overfit these hyperparamters to the data, which can subsequently lead to false-positives. Our findings suggest that the current metrics (SNR and Welch T-test) might overestimate confidences of molecular detections, given that one is allowed to fine-tune detrending hyperparameters. These tests are reasonable for cases in which we directly compare a model with data; however one must be careful when altering the data in order to optimize the detection significance. These tests might be accurate when coupled with a reasonable and consistent detrending method, and without {\it a priori} assumptions about the planetary signal. However, it is difficult to determine detrending hyperparameters without model injection and recovery. We recommend analyses which could help rule out false-positives by: (1) quantifying all detrending hyperparameters; (2) optimizing detrending to a grid of various $K_p$ and $V_{sys}$, including the planetary location; (3) comparing the detection significance at the planetary $K_p$ and $V_{sys}$ to other spurious maxima after each optimization; and (4) reproducing the detections with multiple detrending methods.

Several advancements may help improve the robustness of chemical detections using high-resolution spectroscopy. The accumulation of datasets over multiple nights of observations lead to the strongest, most robust detections \citep{brogi2012}. Ideally, these observations span different orbital phase ranges so that $K_p$ and $V_{sys}$ contours overlap and constrain the planetary signal. Additionally, confidences may improve with wider wavelength coverage, as cross-correlation will involve more spectral features. The upcoming CRIRES+ \citep{follert2014} will provide dramatic improvements upon the existing instrument, including 10$\times$ the wavelength coverage. As discussed in \citet{brogi2017}, high-resolution spectroscopy may also be combined with low-resolution observations to better constrain molecular detections and abundances. Additionally it might be useful to compare observations of identical objects obtained from different spectrographs (e.g. CRIRES, GIANO and NIRSPEC). This might help in distinguishing between spurious peaks due to noise and detection peaks from intrinsic features. The host of existing and upcoming instruments indicates great potential for high-resolution spectroscopy to make new chemical detections, and to constrain compositions of exoplanet atmospheres.

\section*{Acknowledgements}

N.M., G.H., and S.G acknowledge support from the UK Science and Technology Facilities Council (STFC). 
We thank Dr. Matteo Brogi for his insightful comments which helped improve the manuscript. This work is 
based on observations made using the CRIRES spectrograph on the European Southern Observatory 
(ESO) Very Large Telescope (VLT) (186.C-0289). We thank the ESO Science Archive for providing the data. 
This research has made use of NASA's Astrophysics Data System Service. 




\bibliographystyle{mnras}
\bibliography{hires} 

\begin{thebibliography}{}
\makeatletter
\relax
\def\mn@urlcharsother{\let\do\@makeother \do\$\do\&\do\#\do\^\do\_\do\%\do\~}
\def\mn@doi{\begingroup\mn@urlcharsother \@ifnextchar [ {\mn@doi@}
  {\mn@doi@[]}}
\def\mn@doi@[#1]#2{\def\@tempa{#1}\ifx\@tempa\@empty \href
  {http://dx.doi.org/#2} {doi:#2}\else \href {http://dx.doi.org/#2} {#1}\fi
  \endgroup}
\def\mn@eprint#1#2{\mn@eprint@#1:#2::\@nil}
\def\mn@eprint@arXiv#1{\href {http://arxiv.org/abs/#1} {{\tt arXiv:#1}}}
\def\mn@eprint@dblp#1{\href {http://dblp.uni-trier.de/rec/bibtex/#1.xml}
  {dblp:#1}}
\def\mn@eprint@#1:#2:#3:#4\@nil{\def\@tempa {#1}\def\@tempb {#2}\def\@tempc
  {#3}\ifx \@tempc \@empty \let \@tempc \@tempb \let \@tempb \@tempa \fi \ifx
  \@tempb \@empty \def\@tempb {arXiv}\fi \@ifundefined
  {mn@eprint@\@tempb}{\@tempb:\@tempc}{\expandafter \expandafter \csname
  mn@eprint@\@tempb\endcsname \expandafter{\@tempc}}}

\bibitem[\protect\citeauthoryear{{Barber}, {Strange}, {Hill}, {Polyansky},
  {Mellau}, {Yurchenko}  \& {Tennyson}}{{Barber} et~al.}{2014}]{barber_2014}
{Barber} R.~J.,  {Strange} J.~K.,  {Hill} C.,  {Polyansky} O.~L.,  {Mellau}
  G.~C.,  {Yurchenko} S.~N.,   {Tennyson} J.,  2014, \mn@doi [Mon. Not. R.
  Astron. Soc.] {10.1093/mnras/stt2011}, \href
  {http://adsabs.harvard.edu/abs/2014MNRAS.437.1828B} {437, 1828}

\bibitem[\protect\citeauthoryear{{Birkby}}{{Birkby}}{2018}]{birkby2018}
{Birkby} J.~L.,  2018, preprint, \href
  {http://adsabs.harvard.edu/abs/2018arXiv180604617B} {} (\mn@eprint {arXiv}
  {1806.04617})

\bibitem[\protect\citeauthoryear{{Birkby}, {de Kok}, {Brogi}, {de Mooij},
  {Schwarz}, {Albrecht}  \& {Snellen}}{{Birkby} et~al.}{2013}]{birkby2013}
{Birkby} J.~L.,  {de Kok} R.~J.,  {Brogi} M.,  {de Mooij} E.~J.~W.,  {Schwarz}
  H.,  {Albrecht} S.,   {Snellen} I.~A.~G.,  2013, \mn@doi [Mon. Not. R.
  Astron. Soc.] {10.1093/mnrasl/slt107}, \href
  {http://adsabs.harvard.edu/abs/2013MNRAS.436L..35B} {436, L35}

\bibitem[\protect\citeauthoryear{{Birkby}, {de Kok}, {Brogi}, {Schwarz}  \&
  {Snellen}}{{Birkby} et~al.}{2017}]{birkby2017}
{Birkby} J.~L.,  {de Kok} R.~J.,  {Brogi} M.,  {Schwarz} H.,   {Snellen}
  I.~A.~G.,  2017, \mn@doi [Astrophys. J.] {10.3847/1538-3881/aa5c87}, \href
  {http://adsabs.harvard.edu/abs/2017AJ....153..138B} {153, 138}

\bibitem[\protect\citeauthoryear{{Bouchy} et~al.,}{{Bouchy}
  et~al.}{2005}]{bouchy2005}
{Bouchy} F.,  et~al., 2005, \mn@doi [Astron. Astrophys.]
  {10.1051/0004-6361:200500201}, \href
  {http://adsabs.harvard.edu/abs/2005A%26A...444L..15B} {444, L15}

\bibitem[\protect\citeauthoryear{{Brogi}, {Snellen}, {de Kok}, {Albrecht},
  {Birkby}  \& {de Mooij}}{{Brogi} et~al.}{2012}]{brogi2012}
{Brogi} M.,  {Snellen} I.~A.~G.,  {de Kok} R.~J.,  {Albrecht} S.,  {Birkby} J.,
    {de Mooij} E.~J.~W.,  2012, \mn@doi [Nature] {10.1038/nature11161}, \href
  {http://adsabs.harvard.edu/abs/2012Natur.486..502B} {486, 502}

\bibitem[\protect\citeauthoryear{{Brogi}, {Snellen}, {de Kok}, {Albrecht},
  {Birkby}  \& {de Mooij}}{{Brogi} et~al.}{2013}]{brogi2013}
{Brogi} M.,  {Snellen} I.~A.~G.,  {de Kok} R.~J.,  {Albrecht} S.,  {Birkby}
  J.~L.,   {de Mooij} E.~J.~W.,  2013, \mn@doi [Astrophys. J.]
  {10.1088/0004-637X/767/1/27}, \href
  {http://adsabs.harvard.edu/abs/2013ApJ...767...27B} {767, 27}

\bibitem[\protect\citeauthoryear{{Brogi}, {de Kok}, {Birkby}, {Schwarz}  \&
  {Snellen}}{{Brogi} et~al.}{2014}]{brogi2014}
{Brogi} M.,  {de Kok} R.~J.,  {Birkby} J.~L.,  {Schwarz} H.,   {Snellen}
  I.~A.~G.,  2014, \mn@doi [Astron. Astrophys.] {10.1051/0004-6361/201423537},
  \href {http://adsabs.harvard.edu/abs/2014A%26A...565A.124B} {565, A124}

\bibitem[\protect\citeauthoryear{{Brogi}, {de Kok}, {Albrecht}, {Snellen},
  {Birkby}  \& {Schwarz}}{{Brogi} et~al.}{2016}]{brogi2016}
{Brogi} M.,  {de Kok} R.~J.,  {Albrecht} S.,  {Snellen} I.~A.~G.,  {Birkby}
  J.~L.,   {Schwarz} H.,  2016, \mn@doi [Astrophys. J.]
  {10.3847/0004-637X/817/2/106}, \href
  {http://adsabs.harvard.edu/abs/2016ApJ...817..106B} {817, 106}

\bibitem[\protect\citeauthoryear{{Brogi}, {Line}, {Bean}, {D{\'e}sert}  \&
  {Schwarz}}{{Brogi} et~al.}{2017}]{brogi2017}
{Brogi} M.,  {Line} M.,  {Bean} J.,  {D{\'e}sert} J.-M.,   {Schwarz} H.,  2017,
  \mn@doi [Astrophys. J. Lett.] {10.3847/2041-8213/aa6933}, \href
  {http://adsabs.harvard.edu/abs/2017ApJ...839L...2B} {839, L2}

\bibitem[\protect\citeauthoryear{{Brogi}, {Giacobbe}, {Guilluy}, {de Kok},
  {Sozzetti}, {Mancini}  \& {Bonomo}}{{Brogi} et~al.}{2018}]{brogi2018}
{Brogi} M.,  {Giacobbe} P.,  {Guilluy} G.,  {de Kok} R.~J.,  {Sozzetti} A.,
  {Mancini} L.,   {Bonomo} A.~S.,  2018, \mn@doi [Astron. Astrophys.]
  {10.1051/0004-6361/201732189}, \href
  {http://adsabs.harvard.edu/abs/2018A%26A...615A..16B} {615, A16}

\bibitem[\protect\citeauthoryear{{Follert} et~al.,}{{Follert}
  et~al.}{2014}]{follert2014}
{Follert} R.,  et~al., 2014, in Ground-based and Airborne Instrumentation for
  Astronomy V. p. 914719, \mn@doi{10.1117/12.2054197}

\bibitem[\protect\citeauthoryear{{Gandhi} \& {Madhusudhan}}{{Gandhi} \&
  {Madhusudhan}}{2017}]{gandhi_2017}
{Gandhi} S.,  {Madhusudhan} N.,  2017, \mn@doi [Mon. Not. R. Astron. Soc.]
  {10.1093/mnras/stx1601}, \href
  {http://adsabs.harvard.edu/abs/2017MNRAS.472.2334G} {472, 2334}

\bibitem[\protect\citeauthoryear{{Harris}, {Tennyson}, {Kaminsky}, {Pavlenko}
  \& {Jones}}{{Harris} et~al.}{2006}]{harris_2006}
{Harris} G.~J.,  {Tennyson} J.,  {Kaminsky} B.~M.,  {Pavlenko} Y.~V.,   {Jones}
  H.~R.~A.,  2006, \mn@doi [Mon. Not. R. Astron. Soc.]
  {10.1111/j.1365-2966.2005.09960.x}, \href
  {http://adsabs.harvard.edu/abs/2006MNRAS.367..400H} {367, 400}

\bibitem[\protect\citeauthoryear{Hawker, Madhusudhan, Cabot  \& Gandhi}{Hawker
  et~al.}{2018}]{hawker2018}
Hawker G.~A.,  Madhusudhan N.,  Cabot S. H.~C.,   Gandhi S.,  2018, Astrophys.
  J. Lett., 863, L11

\bibitem[\protect\citeauthoryear{{Horne}}{{Horne}}{1986}]{horne1986}
{Horne} K.,  1986, \mn@doi [PASP] {10.1086/131801}, \href
  {http://adsabs.harvard.edu/abs/1986PASP...98..609H} {98, 609}

\bibitem[\protect\citeauthoryear{{Hubeny}}{{Hubeny}}{2017}]{hubeny_2017}
{Hubeny} I.,  2017, \mn@doi [Mon. Not. R. Astron. Soc.] {10.1093/mnras/stx758},
  \href {http://adsabs.harvard.edu/abs/2017MNRAS.469..841H} {469, 841}

\bibitem[\protect\citeauthoryear{{Kaeufl} et~al.,}{{Kaeufl}
  et~al.}{2004}]{kaeufl2004}
{Kaeufl} H.-U.,  et~al., 2004, in {Moorwood} A.~F.~M.,  {Iye} M.,  eds,  Proc.
  of SPIE Vol. 5492, Ground-based Instrumentation for Astronomy. pp 1218--1227,
  \mn@doi{10.1117/12.551480}

\bibitem[\protect\citeauthoryear{{Lockwood}, {Johnson}, {Bender}, {Carr},
  {Barman}, {Richert}  \& {Blake}}{{Lockwood} et~al.}{2014}]{lockwood2014}
{Lockwood} A.~C.,  {Johnson} J.~A.,  {Bender} C.~F.,  {Carr} J.~S.,  {Barman}
  T.,  {Richert} A.~J.~W.,   {Blake} G.~A.,  2014, \mn@doi [Astrophys. J.
  Lett.] {10.1088/2041-8205/783/2/L29}, \href
  {http://adsabs.harvard.edu/abs/2014ApJ...783L..29L} {783, L29}

\bibitem[\protect\citeauthoryear{{Lord}}{{Lord}}{1992}]{lord1992}
{Lord} S.~D.,  1992, Technical report, {A new software tool for computing
  Earth's atmospheric transmission of near- and far-infrared radiation}

\bibitem[\protect\citeauthoryear{{Madhusudhan}}{{Madhusudhan}}{2012}]{madhu2012}
{Madhusudhan} N.,  2012, \mn@doi [Astrophys. J.] {10.1088/0004-637X/758/1/36},
  \href {http://adsabs.harvard.edu/abs/2012ApJ...758...36M} {758, 36}

\bibitem[\protect\citeauthoryear{{Madhusudhan}, {Ag{\'u}ndez}, {Moses}  \&
  {Hu}}{{Madhusudhan} et~al.}{2016}]{madhu2016}
{Madhusudhan} N.,  {Ag{\'u}ndez} M.,  {Moses} J.~I.,   {Hu} Y.,  2016, \mn@doi
  [Space Sci. Rev.] {10.1007/s11214-016-0254-3}, \href
  {http://adsabs.harvard.edu/abs/2016SSRv..205..285M} {205, 285}

\bibitem[\protect\citeauthoryear{{Moses}, {Madhusudhan}, {Visscher}  \&
  {Freedman}}{{Moses} et~al.}{2013}]{moses2013}
{Moses} J.~I.,  {Madhusudhan} N.,  {Visscher} C.,   {Freedman} R.~S.,  2013,
  \mn@doi [Astrophys. J.] {10.1088/0004-637X/763/1/25}, \href
  {http://adsabs.harvard.edu/abs/2013ApJ...763...25M} {763, 25}

\bibitem[\protect\citeauthoryear{{Nugroho}, {Kawahara}, {Masuda}, {Hirano},
  {Kotani}  \& {Tajitsu}}{{Nugroho} et~al.}{2017}]{nugroho2017}
{Nugroho} S.~K.,  {Kawahara} H.,  {Masuda} K.,  {Hirano} T.,  {Kotani} T.,
  {Tajitsu} A.,  2017, \mn@doi [Astron. J.] {10.3847/1538-3881/aa9433}, \href
  {http://adsabs.harvard.edu/abs/2017AJ....154..221N} {154, 221}

\bibitem[\protect\citeauthoryear{{Piskorz} et~al.,}{{Piskorz}
  et~al.}{2016}]{piskorz2016}
{Piskorz} D.,  et~al., 2016, \mn@doi [Astrophys. J.]
  {10.3847/0004-637X/832/2/131}, \href
  {http://adsabs.harvard.edu/abs/2016ApJ...832..131P} {832, 131}

\bibitem[\protect\citeauthoryear{{Richard} et~al.,}{{Richard}
  et~al.}{2012}]{richard_2012}
{Richard} C.,  et~al., 2012, \mn@doi [JQSRT] {10.1016/j.jqsrt.2011.11.004},
  \href {http://adsabs.harvard.edu/abs/2012JQSRT.113.1276R} {113, 1276}

\bibitem[\protect\citeauthoryear{{Rodler}, {Lopez-Morales}  \&
  {Ribas}}{{Rodler} et~al.}{2012}]{rodler2012}
{Rodler} F.,  {Lopez-Morales} M.,   {Ribas} I.,  2012, \mn@doi [Astron.
  Astrophys.] {10.1088/2041-8205/753/1/L25}, \href
  {http://adsabs.harvard.edu/abs/2012ApJ...753L..25R} {753, L25}

\bibitem[\protect\citeauthoryear{{Rodler}, {K{\"u}rster}  \& {Barnes}}{{Rodler}
  et~al.}{2013}]{rodler2013}
{Rodler} F.,  {K{\"u}rster} M.,   {Barnes} J.~R.,  2013, \mn@doi [Mon. Not. R.
  Astron. Soc.] {10.1093/mnras/stt462}, \href
  {http://adsabs.harvard.edu/abs/2013MNRAS.432.1980R} {432, 1980}

\bibitem[\protect\citeauthoryear{{Rothman} et~al.,}{{Rothman}
  et~al.}{2010}]{rothman_2010}
{Rothman} L.~S.,  et~al., 2010, \mn@doi [JQSRT] {10.1016/j.jqsrt.2010.05.001},
  \href {http://adsabs.harvard.edu/abs/2010JQSRT.111.2139R} {111, 2139}

\bibitem[\protect\citeauthoryear{{Schwarz}, {Brogi}, {de Kok}, {Birkby}  \&
  {Snellen}}{{Schwarz} et~al.}{2015}]{schwarz2015}
{Schwarz} H.,  {Brogi} M.,  {de Kok} R.,  {Birkby} J.,   {Snellen} I.,  2015,
  \mn@doi [\aap] {10.1051/0004-6361/201425170}, \href
  {http://adsabs.harvard.edu/abs/2015A%26A...576A.111S} {576, A111}

\bibitem[\protect\citeauthoryear{{Smette} et~al.,}{{Smette}
  et~al.}{2015}]{smette2015}
{Smette} A.,  et~al., 2015, \mn@doi [Astron. Astrophys.]
  {10.1051/0004-6361/201423932}, \href
  {http://adsabs.harvard.edu/abs/2015A%26A...576A..77S} {576, A77}

\bibitem[\protect\citeauthoryear{{Snellen}, {de Kok}, {de Mooij}  \&
  {Albrecht}}{{Snellen} et~al.}{2010}]{snellen2010}
{Snellen} I.~A.~G.,  {de Kok} R.~J.,  {de Mooij} E.~J.~W.,   {Albrecht} S.,
  2010, \mn@doi [Nature] {10.1038/nature09111}, \href
  {http://adsabs.harvard.edu/abs/2010Natur.465.1049S} {465, 1049}

\bibitem[\protect\citeauthoryear{{Snellen}, {de Kok}, {de Mooij}, {Brogi},
  {Nefs}  \& {Albrecht}}{{Snellen} et~al.}{2011}]{2011IAUS..276..208S}
{Snellen} I.,  {de Kok} R.,  {de Mooij} E.,  {Brogi} M.,  {Nefs} B.,
  {Albrecht} S.,  2011, in {Sozzetti} A.,  {Lattanzi} M.~G.,   {Boss} A.~P.,
  eds,  IAU Symposium Vol. 276, The Astrophysics of Planetary Systems:
  Formation, Structure, and Dynamical Evolution. pp 208--211 (\mn@eprint
  {arXiv} {1011.4156}), \mn@doi{10.1017/S1743921311020199}

\bibitem[\protect\citeauthoryear{{Tamuz}, {Mazeh}  \& {Zucker}}{{Tamuz}
  et~al.}{2005}]{tamuz_2005}
{Tamuz} O.,  {Mazeh} T.,   {Zucker} S.,  2005, \mn@doi [Mon. Not. R. Astron.
  Soc.] {10.1111/j.1365-2966.2004.08585.x}, \href
  {http://adsabs.harvard.edu/abs/2005MNRAS.356.1466T} {356, 1466}

\bibitem[\protect\citeauthoryear{{Tennyson} et~al.,}{{Tennyson}
  et~al.}{2016}]{tennyson_2016}
{Tennyson} J.,  et~al., 2016, \mn@doi [Journal of Molecular Spectroscopy]
  {10.1016/j.jms.2016.05.002}, \href
  {http://adsabs.harvard.edu/abs/2016JMoSp.327...73T} {327, 73}

\bibitem[\protect\citeauthoryear{{Triaud} et~al.,}{{Triaud}
  et~al.}{2009}]{triaud2009}
{Triaud} A.~H.~M.~J.,  et~al., 2009, \mn@doi [Astron. Astrophys.]
  {10.1051/0004-6361/200911897}, \href
  {http://adsabs.harvard.edu/abs/2009A%26A...506..377T} {506, 377}

\bibitem[\protect\citeauthoryear{{de Kok}, {Brogi}, {Snellen}, {Birkby},
  {Albrecht}  \& {de Mooij}}{{de Kok} et~al.}{2013}]{dekok2013}
{de Kok} R.~J.,  {Brogi} M.,  {Snellen} I.~A.~G.,  {Birkby} J.,  {Albrecht} S.,
    {de Mooij} E.~J.~W.,  2013, \mn@doi [Astron. Astrophys.]
  {10.1051/0004-6361/201321381}, \href
  {http://adsabs.harvard.edu/abs/2013A%26A...554A..82D} {554, A82}

\makeatother
\end{thebibliography}






\bsp	
\label{lastpage}
\end{document}